\documentclass[32pt, usenatbib]{mn2e}
 \usepackage{graphicx,times,subfigure}
\usepackage{bm}
\usepackage{epsf}
\usepackage{amsmath}
\usepackage{amssymb}
\usepackage{cases}
\usepackage{float}
\usepackage{colortbl}
\usepackage{color}
\usepackage{multirow}
\def\la{\langle}

\def\n{\noindent}
\def\be{\begin{equation}}
\def\ee{\end{equation}}
\def\ben{\begin{eqnarray}}
\def\een{\end{eqnarray}}

\def\myC{{\cal C}}

\def\y8y{{}_sY^*_{lm}(\thet,\phi)}

\def\cb1{\textcolor{blue}}

\def\cb{\textcolor{black}}
\begin{document}
\onecolumn
\title[Probing Modified Gravity Theories]
{Galaxy Clustering in 3D and Modified Gravity Theories}
\author[Munshi et al.]
{D. Munshi$^{1}$, G. Pratten$^{1}$,  P. Valageas$^{2,3}$, P. Coles$^{1}$, P. Brax$^{2,3}$ \\
$^{1}$ Astronomy Centre, School of Mathematical and Physical Sciences, University of Sussex, Brighton BN1 9QH, United Kingdom\\
$^{2}$ CEA, IPhT, F-91191, Gif-sur-Yevette, Ce\'dex, France\\
$^{3}$ CNRS, URA, 2306, F-91191, Gif-sur-Yevette, Ce\'dex, France}
\maketitle
\begin{abstract}
We study Modified Gravity (MG) theories by modelling the redshifted 
matter power spectrum in a spherical Fourier-Bessel (sFB) basis. 
We use a fully non-linear description of the real-space matter power-spectrum
and include the lowest-order redshift-space correction (Kaiser effect), taking into account
some additional non-linear contributions.
Ignoring relativistic corrections, which are not expected to play an important role 
for a shallow survey, we analyse two different modified gravity scenarios, namely the 
{\em generalised} Dilaton scalar-tensor theories and the
$f({R})$ models in the large curvature regime. 
We compute the 3D power spectrum ${\cal C}^s_{\ell}(k_1,k_2)$
for various such MG theories with and without redshift space distortions, assuming  
precise knowledge of background cosmological parameters. 
Using an all-sky {\em spectroscopic} survey with Gaussian selection function 
$\varphi(r)\propto \exp(-{r^2/r^2_0}), r_0=150h^{-1}${\rm Mpc}, and number density of
galaxies $\bar {\rm N} =10^{-4}\;{\rm Mpc}^{-3}$, we use a $\chi^2$ analysis, and find 
that the lower-order $(\ell \leq 25)$ multipoles of ${\cal C}^s_\ell(k,k')$ (with radial modes restricted to 
$k<0.2 h {\rm Mpc}^{-1}$) can constraint the parameter $f_{R_0}$ at a level of 
$2\times 10^{-5} (3\times 10^{-5})$ with $3 \sigma$ confidence for $n=1(2)$. 
Combining constraints from higher $\ell>25$ modes can further reduce the error bars 
and thus in principle make cosmological gravity constraints competitive with solar system 
tests. However this will require an accurate modelling of non-linear redshift 
space distortions.
Using a {\em tomographic} $\beta(a)$-$m(a)$ parameterization
we also derive constraints on specific parameters describing the Dilaton models of modified gravity. 
\end{abstract}
\begin{keywords}: Cosmology-- Modified Gravity Theories -- Methods: analytical, statistical, numerical
\end{keywords}
\section{Introduction}
The apparent accelerated expansion of the Universe \citep{Perl99, Ries98} can be explained within General Relativity (GR)
by introducing a finely tuned cosmological constant. However, there are alternative explanations for this phenomenon,
including modified gravity theories. While the laws of gravity are not well constrained 
on cosmological scales \citep{JJKT14, Clift12}, modification of GR are tightly constrained in the solar 
system \citep{OS03} or at astrophysical scales \citep{JVS13,VCJV13}.

As shown by various authors \citep{Bertschinger:2006aw,SHS07,Brax:2008hh} the background dynamics in various dark energy 
and modified gravity models are 
nearly indistinguishable. Thus it is important to investigate the evolution of perturbations in these models. 
The studies of perturbation theory in modified gravity models, in principle, can be classified in two different frameworks: 
the parametric approach and the non-parametric method,  
e.g. the principal component analysis~\citep{Zhao:2008bn,Zhao:2009fn,Zhao:2010dz,Hojjati:2011xd,Hall13}. 
Several parametrizations of modified gravity have been proposed for the evolution of linear perturbations.
Such parametrizations typically involve two functions $\nu(k,a)$ and $\gamma(k,a)$ that both depend
on the scale factor $a$ and wave number $k$, when they describe the modification to
the Poisson equation for the metric potentials. We will use instead the recently proposed parametrization of the coupling to matter $\beta(a)$ and the mass $m(a)$ of the scalar field, which only depend on the scale factor (but give rise to both time and scale dependences in the resulting modification to the Poisson equation). 
The $\beta(a)-m(a)$ offers a unified approach to study various modified gravity models including
$f(R)$ gravity and the symmetron and Dilaton models \citep{BDL12}. Parametrization based on effective field theory (EFT)
has also been considered \citep{Cheung08}.
%
Besides the recent progress on the construction of parametrizations, many observational windows have recently been proposed, such as the galaxy clustering \citep{PS08,OLH}, Integrated Sachs-Wolfe (ISW) effect in Cosmic Microwave Background (CMB) anisotropies~\citep{Zhang06}, the galaxy-ISW cross correlation~\citep{Song:2007da}, cluster abundance~\citep{Jain:2007yk,Lombriser:2010mp}, peculiar velocity~\citep{Li:2012by}, redshift-space distortions~\citep{Jennings:2012pt,Guzz08}, weak-lensing~\citep{HKV07,Sch08,TT08}, $21$cm observations~\citep{Hall13}, matter bispectrum~\citep{GilMarin:2011xq}, {\it etc}. In addition, recently some N-body simulation algorithms in modified gravity models have been developed~\citep{Zhao:2010qy,Li:2010zw}. As shown by various authors \citep{Song:2007da,Lombriser:2010mp}, at the WMAP resolution the modification effects on the CMB mainly come from the ISW effect, which becomes prominent on the super-horizon scales. However, due to the unavoidable cosmic variance on large scales, the constraints from these effects are not significant. On the other hand, since the typical modification scales are in the sub-horizon range, several studies show that the most stringent cosmological constraints come from the large-scale structure data sets. 
For instance, in the case of $f(R)$ gravity one obtains 
$|f_{R_0}| \leq 6.5 \times 10^{-5}\;,95\%{\rm C.L.}$ ~\citep{DHP14} from the combined
analysis of the CMB temperature power spectrum, the galaxy power spectrum
and the baryon acoustic oscillations measurements, or 
$|f_{R_0}| \leq 4.6 \times 10^{-5}\;,95\%{\rm C.L.}$ ~\citep{Julien14} from the
analysis of the galaxy power spectrum through the clustering ratio.
This is competitive with solar system constraints but astrophysical constraints
in dwarf galaxies provide tighter bounds $|f_{R_0}| \leq 5 \times 10^{-7}$
~\citep{JVS13,VCJV13}

The past few decades have seen a rapid progress in large scale galaxy surveys.
SDSS\footnote{http://www.sdss.org/} and 2dFGSS\footnote{http://www.roe.ac.uk/~jap/2df/} opened a 
new horizon in modern cosmology by mapping three-dimensional positions of millions of galaxy.
The BOSS and DES represent current state-of-the art galaxy surveys, together with the recently completed  
WiggleZ\footnote{http://wigglez.swin.edu.au/site/}. 
Future surveys such as the Euclid\footnote{http://sci.esa.int/euclid}\citep{Laure11}, LSST\footnote{http://www.lsst.org/lsst/} 
and Wide-Field InfraRed Survey Telescope\footnote{http://wfirst.gsfc.nasa.gov/}
will measure galaxy clustering with greatly increased statistical power.
They will test the theory of General Relativity (GR) on cosmological scales.
One way to do so is to examine the growth of structure using the $\gamma$-parameter (to be introduced later). 
Previous calculations suggest that Euclid can constrain $\gamma$ to $0.01$ (for $\Lambda$CDM  $\gamma$ corresponds to $\gamma \simeq 0.55$) \citep{Linder05, HKV07}.
Generic modification of gravity may require more than just one parameter \citep{AKS08, FS10} but previous studies have 
shown that a Euclid-type survey could measure these parameters to high
precision \citep{Dan10, Amend10}.

The power spectrum of density inhomogeneities in the nearby Universe and the temperature fluctuations in the cosmic microwave background (CMB) sky carry the bulk of the cosmological information and are routinely employed in analysing data from CMB experiments as well as large-scale structure (LSS) surveys.They carry complementary cosmological information; all-sky CMB observations such as NASA's WMAP\footnote {http://map.gsfc.nasa.gov/}and ESA's Planck\footnote{http://www.rssd.esa.int/planck} experiments primarily  probe the distribution of matter and radiation at redshift $z=1300$. Such surveys are projected or 2D surveys. 
The large-scale surveys such as those obtained with ESA's Euclid mission will give us a window at a lower redshift range $z \approx 0-2$ and are designed to provide a three-dimensional (3D) map of the Universe using more than fifty million galaxies with spectroscopic redshifts. There are other LSS surveys that are either being planned or in various stages of developments e.g. \citep{AnnisBridle05} for Dark Energy Survey (DES)\footnote{http://www.darkenergysurvey.org/}\citep{Tyson04} for Large Synoptic Survey Telescope (LSST) and \citep{Sch07} for Baryon Oscillation Spectroscopic Survey (BOSS)\footnote{http://cosmology.lbl.gov/BOSS/}. These surveys are designed to map the local dark Universe using weak-lensing of galaxies as well as studying baryonic oscillation features in the matter power spectrum. This has motivated in recent years a flurry of activity in developing 3D power spectrum analysis for cosmological data using spherical Fourier-Bessel (sFB) decomposition (see e.g. \cite{HT95,CHK05,Erd06a,Ab10,Shapiro11,Rassat11,LRS11,Asorey12,PM13}).The sFB power spectrum recovers the 3D power spectrum at each wave-number $k$ and its angular dependence is encoded in the angular momentum $\ell$.

In this paper we use the results derived in \citep{PM13}, to derive the 3D density power
spectrum for models of modified gravity theories. The primary goal is to check to what
extent such power spectrum analysis can be used to constrain departure from GR
using future redshift surveys.

This paper is organised as follows: in \textsection{\ref{sec:theory}} we review different types of modified gravity
theories with special attention to $f(R)$ gravity theories as well as the models known as the {\em generalised} Dilaton models.
We introduce the 3D power-spectrum in \textsection{\ref{sec:red}} using a sFB transform, which provides a 
natural framework for treating redshift space distortions. 
For the specific  baseline $\Lambda$CDM  cosmology we take the following parameter values:
$h=0.73, \Omega_{\rm M}=0.24,  \Omega_{\rm DE}=0.76, \Omega_{\rm K}=0, w_{\rm DE}=-1, \sigma_8=0.76, n_s=0.958$. 
\section{Modified Gravity Theories}
\label{sec:theory}
Henceforth, we consider modified-gravity theories that differ from the standard $\Lambda$-CDM
cosmology by adding a new degree of freedom, which can be associated to a scalar field $\varphi$.
This new component can drive the accelerated expansion of the Universe at late times,
typically through the nonzero value of the minimum of its potential, although this can be seen as
introducing a simple cosmological constant. The main difference from the $\Lambda$-CDM cosmology
and quintessence scenarios, which only modify the background expansion, is that the fifth force induced
by the scalar field also modifies the growth of structures, typically giving a new scale dependence to the
linear growing modes of density perturbations and accelerating the collapse of overdense regions.

These modified-gravity theories can be classified in three broad categories:
{\it Chameleon}, {\it K-mouflage} and {\it Vainshtein} scenarios,
according to their nonlinear
screening mechanism that ensures convergence to General Relativity in small-scale and high-density 
environments such as the Solar System (as the high-accuracy Solar-System
measurements provide tight constraints on local deviations from General Relativity).
The K-mouflage and Vainshtein mechanisms rely on non-standard kinetic 
terms that drive spatial gradients of the scalar field to zero in high-density
regions and suppress the fifth force.
Chameleon scenarios typically contain additional couplings between the
scalar field and the metric, or new geometric terms beyond the Einstein-Hilbert gravitational action. These two equivalent descriptions can be captured by the $\{\beta(a), m(a)\}$ parameterisation which will be used throughout this work 
\citep{BDL12,BDLW12}. 

In this paper, we focus on two different models of the Chameleon type.
In the {\it Dilaton} models, the coupling between the scalar field
and the metric depends on the scalar field value, so that in
dense environments where the scalar field is driven to zero, the coupling
also vanishes, which suppresses the fifth force as in the 
Damour-Polyakov mechanism \citep{DP94}.
In the $f(R)$ models, the coupling of the scalar field to matter is constant
but its effective potential depends on the environment and its mass
becomes large in high-density regions. This suppresses
the magnitude of the fifth force through a Yukawa screening.
In all chameleon cases, these modifications of gravity induce a global enhancement of the effective force of gravity, due to the fifth force,
which directly translates into an increase of structure formation.
In this Section, we review the {\it Dilaton} 
and the $f(R)$ models and we describe parameterization in the context of large-scale structure formation.
\subsection{Gravity in Dilaton models}
\label{sec:Dilatons}

\begin{table}
\begin{center}
\caption{Parameters describing the Dilaton Models considered in our study. The parameters are used to define the scalar potential
$V(\varphi)$ and the coupling function $A(\varphi)$ through the $\{\beta(a),m(a)\}$
parameterization defined in Eqs.(\ref{eq:beta-a-def})-(\ref{eq:s-def}).}
\begin{tabular}{| c | c | c | c | c | }
  \hline
  Model & $m_0 \;\; [h {\rm Mpc}^{-1}]$ & $r$ & $\beta_0$  & $s$ \\
  \hline
  \rowcolor[gray]{0.8}  (A1,A2,A3) & $(0.334,0.334,0.334)$ & $(1.0,1.0,1.0)$ & $(0.5,0.5,0.5)$ & $(0.6,0.24,0.12)$ \\
  \hline
  (B1,B3,B4) & $(0.334,0.334,0.334)$ & $(1.0,1.0,1.0)$ & $(0.25,0.75,1.0)$ & $(0.24,0.24,0.24)$ \\
  \hline
  \rowcolor[gray]{0.8} (C1,C3,C4) & $(0.334,0.334,0.334)$ & $(1.33,0.67,0.4)$ & $(0.5,0.5,0.5)$ & $(0.24,0.24,0.24)$ \\
  \hline
  (E1,E3,E4) & $(0.667,0.167,0.111)$ & $(1.0,1.0,1.0)$ & $(0.5,0.5,0.5)$ & $(0.24,0.24,0.24)$ \\
  \hline
\end{tabular}
\label{tabular:tab2}
\end{center}
\end{table}

The {\it Dilaton} theories of modified gravity are chameleon models with the
the {\it Damour-Polyakov} property \citep{DP94}, according to which the coupling between the scalar field $\varphi$ and the rest of the matter components approaches zero in
dense environments \citep{Piet05, OP08, HK10}.
In contrast with the $f(R)$ theories described in Sec.~\ref{sec:f(R)} below,
the scalar field here takes on a small mass everywhere and thus mediates a long-range (screened) force.
These Dilaton models are scalar-tensor theories, where the action defining the system
takes the general form
\ben
&& S=\int d^4x \sqrt{-g} \left[ \frac{M_{\rm Pl}^2}{2} R - \frac{1}{2} (\nabla\varphi)^2 
- V(\varphi) - \Lambda_0^4 \right] + \int d^4x \sqrt{-\tilde{g}} \tilde{\cal L}_m( \psi^{(i)}_m,\tilde{g}_{\mu\nu}) ,
\label{eq:S-dilaton-def}
\een
where $M_{\rm Pl}=(8\pi {\cal G}_{\rm N})^{-1/2}$ is the reduced Planck mass (in natural units),
$g$ is the determinant of the Einstein-frame metric tensor $g_{\mu\nu}$ and
$\tilde{g}$ the determinant of the Jordan-frame metric tensor $\tilde{g}_{\mu\nu}$,
which is given by the conformal rescaling
\ben
&& \tilde{g}_{\mu\nu} = A^2(\varphi) g_{\mu\nu} .
\label{conformal}
\een
The various matter fields $\psi^{(i)}_m$ are governed by the Jordan-frame
Lagrangian density $\tilde{\cal L}_m$ and the scalar field $\varphi$ by the
Einstein-frame Lagrangian density ${\cal L}_{\varphi}=-1/2(\nabla\varphi)^2-V(\varphi)$,
with the scalar-field potential $V(\varphi)$.
There is no explicit coupling between matter and the scalar field and the fifth
force on matter particles due to the scalar field arises from the conformal
transformation (\ref{conformal}) (more precisely, through gradients of $A$).
In the Lagrangian presented in Eq.(\ref{eq:S-dilaton-def}) we explicitly wrote the cosmological
constant term $\Lambda_0^4$, so that the minimum of $V(\varphi)$ is zero,
which is reached for $\varphi \rightarrow \infty$, but this could also be interpreted
as the non-zero minimum of the scalar field potential.

In the original Dilaton model, the potential of the scalar field $V(\varphi)$ and its 
coupling $A(\varphi)$ with the metric have the following functional forms:
\ben
&& V(\varphi) = V_0 \exp \left( - {\varphi \over M_{\rm Pl}} \right) ,
\label{V-exp-dilaton} \\
&& A(\varphi) = 1 + \frac{A_2}{2} \frac{\varphi^2}{M^2_{\rm Pl}} ,
\label{A-phi-def}
\een
where $\{V_0, A_2 \}$ are the two free parameters.
In dense regions where  $\varphi \approx 0$, the coupling to matter is negligible,
and gravity converges to GR.
However, the field nevertheless mediates a long range gravitational force that has an effect  elsewhere, in less dense environments.
This model can be generalized to a greater class of Dilaton models, by keeping the coupling
function as in Eq.(\ref{A-phi-def}) but considering more general potentials.
Then, instead of specifying the model by its potential $V(\varphi)$ it is convenient
to define the model by the tomographic parametrization $\{ \beta(a), m(a) \}$ 
\citep{BDL12, BV13}, in terms of the scale factor $a(t)$,
where the coupling $\beta(a)$ and the scalar field mass $m(a)$ are defined as
\ben
&& \beta(a) \equiv \beta[\bar{\varphi}(a)] = M_{\rm Pl} \frac{d\ln A}{d\varphi}(\bar{\varphi}) , 
\label{eq:beta-a-def} \\
&& m^2(a) \equiv m^2[\bar{\varphi}(a),\bar{\rho}(a)] = \frac{1}{c^2} \left[
\frac{d^2V}{d\varphi^2}(\bar{\varphi}) + \bar{\rho} \frac{d^2 A}{d\varphi^2}(\bar{\varphi}) 
\right] .
\label{eq:m-a-def}
\een
In this paper we consider the simple forms
\ben
&& m(a)=m_0 \, a^{-r} , \;\;\; \beta(a)=\beta_0 \exp\left[-s \frac{a^{2r-3}-1}{3-2r} \right] ,
\label{eq:symm}
\een
with
\ben
&& s = \frac{9 A_2 \Omega_{m0}H_0^2}{c^2 m_0^2 } .
\label{eq:s-def}
\een
[The exponential potential Eq.(\ref{V-exp-dilaton}) corresponds to $r=3/2$.]
The values of the free parameters $\{m_0, r, \beta_0, s\}$ that enter Eq.(\ref{eq:symm})  
are displayed in Table \ref{tabular:tab2}.
The models \{A,B,C\} were chosen such as to correspond to those studied in 
\citet{BV13} and \citet{BDLWZ12}, where detailed comparisons between numerical 
and analytical calculations are presented.
The models \{A,B,C,E\} probe the dependence on $\{s,\beta_0,r,m_0\}$ respectively,
other parameters being fixed (instead of the models ``D'' considered in 
\citet{BV13}, which probe the dependence on $m_0$ at fixed $A_2$,
we introduced the models E that probe the dependence on the parameter $m_0$
at fixed $\{s,\beta_0,r\}$).
These models probe deviations from the LCDM cosmology of less than $20\%$, in terms
of the matter power spectrum.

In these Dilaton models, the coupling function $A$ is always very close to unity,
so that most Einstein-frame and Jordan-frame quantities (e.g., Hubble expansion rates
or densities) are almost identical.
Indeed, from Eqs.(\ref{A-phi-def}), (\ref{eq:beta-a-def}) 
and (\ref{eq:s-def}) we obtain $\bar{A} \simeq 1 + \beta^2/(2 A_2)$
and $A_2 \sim (c m_0/H_0)^2$.
Solar System tests of gravity imply that $m_0 \gtrsim 10^3 H_0/c$, whence
$A_2 \gtrsim 10^6$, $H/c m \lesssim 10^{-3}$ and $| \bar{A} - 1 | \lesssim 10^{-6}$.
Therefore, the Jordan-frame and Einstein-frame scale factors and background matter 
densities, related by $\tilde{a} = \bar{A} a$ and $\tilde{\bar\rho} = \bar{A}^{-4} \bar{\rho}$,
can be considered equal, as well as the cosmic times and Hubble expansion rates. 
(However, in this section we work in the Einstein frame, where the analysis of the gravitational
dynamics is simpler.)

In the Einstein frame, the Friedmann equation takes the usual form,
\ben
&& 3 M_{\rm Pl}^2 H^2 = \bar{\rho} + \bar{\rho}_{\varphi} + \bar{\rho}_{\Lambda} ,
\label{eq:Friedmann-dilaton}
\een
where we consider the matter and scalar field components and the cosmological
constant contribution $\bar{\rho}_{\Lambda}$.
One can check that the scalar field energy density is negligible as compared with
the matter density, $\bar\rho_{\varphi}/\bar\rho \sim 10^{-6}$, so that the Friedmann equation (\ref{eq:Friedmann-dilaton}) is governed by the matter
density and the cosmological constant and we recover the $\Lambda$-CDM cosmological
expansion, $3 M_{\rm Pl}^2 H^2 = \bar{\rho} + \bar{\rho}_{\Lambda}$,
up to an accuracy of $10^{-6}$.
In the Newtonian gauge, the perturbed metric can be written as
\begin{eqnarray}
&& ds^2 = - (1 + 2\Phi) dt^2 + a^2(t) (1-2\Psi) \delta_{ij}dx^idx^j ,
\label{eq:perturbed-metric}
\end{eqnarray}
where $\Phi$ and $\Psi$ are the Einstein-frame metric gravitational potentials.
One can check that the impact of the scalar field fluctuations on the metric potentials
are again negligible, as $|\delta\rho_{\varphi}|/|\delta\rho| \lesssim 10^{-6}$,
and we have within a $10^{-6}$ accuracy
$\Phi = \Psi = \Psi_{\rm N}$,
where $\Psi_{\rm N}$ is the Newtonian potential given by the Poisson equation,
\begin{eqnarray}
&& \frac{\nabla^2}{a^2} \Psi_{\rm N} = 4\pi {\cal G}_{\rm N} \delta\rho
= \frac{3 \Omega_{m0} H_0^2}{2 a^3} \delta ,
\label{eq:Poisson-GR}
\end{eqnarray}
where $\delta=\delta\rho/\bar{\rho}$ is the matter density contrast.
However, the dynamics of matter particles is modified by the scalar field, which 
gives rise to a fifth force given by ${\bf F}_{A} = - c^2 \nabla \ln A$, that is, in the
Euler equation we must add to the Newtonian potential $\Psi_{\rm N}$ a
fifth-force potential $\Psi_A=c^2 \ln A$ that is not negligible 
and can lead to $10\%$ deviations to the matter density power spectrum
for the parameters given in Table~\ref{tabular:tab2}
(indeed, whereas $|A-1| \lesssim 10^{-6}$ is negligible as compared with unity,
it is not negligible as compared with $|\Psi_{\rm N}|/c^2 \lesssim 10^{-5}$).
\subsection{{Gravity in $f(R)$ theories}}
\label{sec:f(R)}
In models known as the $f(R)$ gravity, the Einstein-Hilbert action $S_{\rm GR}$
is modified by promoting the Ricci scalar $R$ to a function of $R$  \citep{Buch70,Staro80,Staro07,HS07}.
The new action $S$ for the $f(R)$ gravity theories can be written as:
\begin{eqnarray}
&& S = \int d^4 x \sqrt{-g} \, \left[ \frac{M^2_{\rm Pl}}{2} [ R + f(R) ] - \Lambda_0^4  
+ {\cal L}_m(\psi_m^{(i)}) \right] ,
\label{S-fR-def}
\end{eqnarray}
where we explicitly wrote the cosmological constant contribution $\Lambda_0^4$,
although it is often included within the function $f(R)$ [with our choice $f(R)$ describes
the deviations from GR and from the $\Lambda$CDM cosmology].
In this section, contrary to the previous section \ref{sec:Dilatons} where we studied Dilaton
models, we denote with a tilde Einstein-frame quantities instead of Jordan-frame ones, 
because we now work in the Jordan frame.
In the parameterization of \citet{HS07}, the functional form $f({R})$ can be expressed in the high curvature limit as
\begin{eqnarray}
&& f({R}) =  - {f_{R_0} \over n}{{R}_0^{n+1} \over {R}^n}, 
\;\;\; f_R \equiv {df(R) \over dR} = f_{R_0} \frac{R_0^{n+1}}{R^{n+1}}.
\label{eq:fofR}
\end{eqnarray}
The two independent parameters, $f_{R_0}<0$ and $n>0$, can be constrained by observations.
In the above expression, $R_0$ is the present value of the  Ricci scalar for the cosmological background.
Note that this parametrization and that of \citet{Staro07} both reproduce the same results in the large curvature regime.

The $f(R)$ theories of gravity invoke the {\it Chameleon} mechanism to screen modifications of GR in dense environments such as in our Solar System.
In this model, this occurs by requiring that all extra terms vanish in high curvature environment, i.e by requiring $f(|R| \gg |R_0|) \rightarrow 0$.
In such a theory, the background expansion follows the $\Lambda$-CDM dynamics
(as observational constraints imply $|f_{R_0}|\ll 1$),
and the growth of structure is only affected on intermediate and quasi-linear scales.

There is an essential connection between the formulation of the $f(R)$ theory presented above, 
and scalar-tensor  theories of modified gravity.
Upon the coordinate rescaling $\tilde  g_{\mu\nu}= A^{-2}(\varphi)g_{\mu\nu}$
(recall that in this section $\tilde{g}_{\mu\nu}$ is the Einstein-frame metric), 
with $A(\varphi)= \exp[\beta \varphi/ M_{\rm Pl}]$ and $\beta = {1/ \sqrt 6}$,
the $f(R)$ modifications to GR are re-cast as arising from contributions of an extra scalar field $\varphi$, that is subject to a potential $V(\varphi)$ given by:
\begin{eqnarray}
&& V(\varphi) = \frac{M^2_{\rm Pl}}{2} \left( \frac{R f_R -f(R)}{(1+f_R)^2} \right) ,
\;\;\; f_R= \exp \left [ -\frac{2\beta\varphi}{M_{\rm Pl}} \right ] - 1 .
\end{eqnarray}
In that sense,  $f(R)$ theories are equivalent to a scalar-tensor theory expressed in the  Einstein frame \citep{Chiba03,NS04}.
In this new formulation,  the screening mechanism takes another form:
the mass of the scalar field grows with matter density, and a Yukawa-like potential suppresses the fifth force in dense environments. This can be conveniently reformulated by saying that
screening takes place wherever the scalar field  is small compared to the ambient Newtonian potential.

It turns out that all chameleon-like models such as $f(R)$ theories can again be parameterised by the value of the mass $m(a)$ and the coupling $\beta (a)$ of the scalar field, in terms 
of the scale factor $a$ and the associated background matter density 
$\bar{\rho}(a)= {3 \Omega_{m0} H_0^2 M_{\rm Pl}^2}/{a^3}$.
With the specific functional form of $f(R)$ given by Eq.(\ref{eq:fofR}), we can  directly 
relate  $\{n , f_{R_0} \}$ to $\{\beta(a), m(a) \}$ via:
\begin{eqnarray}
&& m(a)=m_0  \left( \frac{4\Omega_{\Lambda 0} +\Omega_{m0} a^{-3}}
{4\Omega_{\Lambda 0} +\Omega_{m0}} \right)^{(n+2)/2}, \;\;\;
m_0 = \frac{H_0}{c} \sqrt{ \Omega_{\rm m0} + 4 \Omega_{\Lambda 0}\over (n+1) | f_{R0} |  } , \;\;\; 
\beta(a) = \frac{1}{\sqrt{6}} . 
\label{m-beta-fR}
\end{eqnarray}
In this paper, we consider values of  $n=\{1,2\}$ and $|f_{R0}| = \{10^{-4},10^{-5},10^{-6}\}$.
The larger value of $|f_{R0}|$ is currently  ruled out by other independent  probes, so this serves as  a consistency test.

As for the Dilaton models described in Sec.~\ref{sec:Dilatons}, the $f(R)$ models
that we consider in this paper follow very closely the $\Lambda$-CDM cosmology
at the background level, because $|f_{R_0}| \ll 1$.
Indeed, from the action (\ref{S-fR-def}) one obtains the Friedmann equation as
\ben
&& 3 M_{\rm Pl}^2 \left[ H^2 - \bar{f}_R (H^2+\dot{H}) + \bar{f}/6 
+ \bar{f}_{RR} H \dot{\bar{R}} \right] =  \bar{\rho} + \bar{\rho}_{\Lambda} ,
\label{Friedmann-fR}
\een
where the dot denotes the derivative with respect to cosmic time $t$ and
$f_{RR} =d^2f/dR^2$. In the background we have $\bar{R} = 12 H^2+6 \dot{H}$
and we can check that all extra terms in the brackets in Eq.(\ref{Friedmann-fR}) 
are of order $|f_{R_0}| H^2$, so that we recover the $\Lambda$-CDM expansion,
$3 M_{\rm Pl}^2 H^2 = \bar{\rho} + \bar{\rho}_{\Lambda}$,
up to an accuracy of $10^{-4}$ for $|f_{R_0}| \lesssim 10^{-4}$.
Moreover, the conformal factor $A(\varphi)$ is given by $A=(1+f_R)^{-1/2}$,
so that $|\bar{A}-1| \lesssim 10^{-4}$ and the background quantities associated with
the Einstein and Jordan frames can be considered equal, and equal to the
$\Lambda$-CDM reference, up to an accuracy of $10^{-4}$.
Considering the metric and density perturbations, we can again write the 
Newtonian gauge metric as in Eq.(\ref{eq:perturbed-metric}) (but this is now the
Jordan-frame metric). Then, in the small-scale sub-horizon limit, 
the modified Einstein equations lead to
\ben
&&\Phi = \Psi_{\rm N} - \frac{c^2}{2} \delta f_R , \;\;\;
\Psi = \Psi_{\rm N} + \frac{c^2}{2} \delta f_R ,
\label{Phi-Psi-fR}
\een
where $\delta f_R= f_R - \bar{f}_F$ and $\Psi_{\rm N}$ is the Newtonian gravitational 
potential defined as in GR by Eq.(\ref{eq:Poisson-GR}).
Thus, because we work in the Jordan frame, in contrast with the Dilaton case
presented in Sec.~\textsection\ref{sec:Dilatons}, the modification of gravity directly appears
through the metric potentials Eq.(\ref{Phi-Psi-fR}).
Finally, the dynamics of the matter particles is given by the geodesic equation,
or the Euler equation in the large-scale single-stream limit, where the Newtonian
potential that appears in GR is replaced by the potential $\Phi$ given in
Eq.(\ref{Phi-Psi-fR}).
\subsection{Modified gravity and structure formation}
\label{sec:Pk_mod_grav}
\subsubsection{Impact of modified gravity on 3D matter clustering}
As we have seen in the previous sections, Dilatons and $f(R)$ theories 
reproduce the smooth background expansion history of the 
standard $\Lambda$-CDM cosmology (up to an accuracy of $10^{-4}$
or better that is sufficient for our purposes).
To distinguish between, and hopefully test such competeting gravitational theories
it is thus necessary to analyse the evolution of matter or metric 
perturbations. To lowest order in cosmological
perturbations, non-standard gravitational scenarios effectively result in a time- and 
scale-dependent modification of the Newtonian constant ${\cal G}_{\rm N}$. 
In the nonlinear regime,
the modifications become more complex as they become sensitive
to the screening mechanism, which depends nonlinearly on the
environment and modifies the form of the equations of motion
(e.g., the effective Poisson equation is no longer linear).
These modifications induce a distortion of the dynamical as well as statistical properties 
of the matter clustering statistics.
To this end, we focus on the signature of these effects on the
matter density power spectrum $P(k,z)$, or more precisely its expansion
${\cal C}_{\ell}(k)$ on spherical multipoles, which is well suited to the cosmological
analysis of wide galaxy surveys.
(We also include redshift-space distortions through a simple approximation
that is exact at linear order and includes some non-linear contributions.)

Thus, we need a way of modelling of the matter power spectrum that applies
to the standard $\Lambda$-CDM scenario as well as these modified-gravity
models.
We use the approach developed in \citep{VNT13}, for the $\Lambda$-CDM 
cosmology, and next extended to these modified-gravity cosmologies
in \citep{BV13}.
This method combines the results from one-loop perturbation theory with that from halo 
model predictions. By construction, this power spectrum agrees with perturbation theory
up to order $P_{L}^2$. In the $\Lambda$-CDM cosmology, this corresponds to the
standard one-loop diagrams \cite{Bernardeau02}. In the modified gravity models,
the linear propagators and the vertices are modified, with new scale and time
dependences. In addition, the effective Poisson equation, obtained after integrating
over the scalar field, becomes nonlinear. At one-loop order this gives rise to a new
diagram for the power spectrum. Beyond the perturbative regime, predictions from a 
suitably altered halo model are incorporated in the high-$k$ limit. The impact of
the modified gravity in the nonlinear dynamics is taken into account
through the halo mass function but the impact of the modified gravity
on the halo profiles is ignored, i.e., we keep the NFW profile from \cite{NFW96}
and the mass-concentration relation from \cite{VNT13}.
The resulting matter density power spectrum $P_{\delta\delta}(k)$ has been tested 
against numerical simulations and found to be in agreement for the
entire available $k$ range from simulations $k< 3h^{-1} \rm Mpc$
\citep{BV13}. 
\subsubsection{Numerical results for the linear and non-linear 3D power spectra}
\begin{figure}
\centering
\vspace{0.15cm}
\begin{tabular}{ccc}
{\epsfxsize=5.25 cm \epsfysize=5.25 cm{\epsfbox[21 21 736 655]{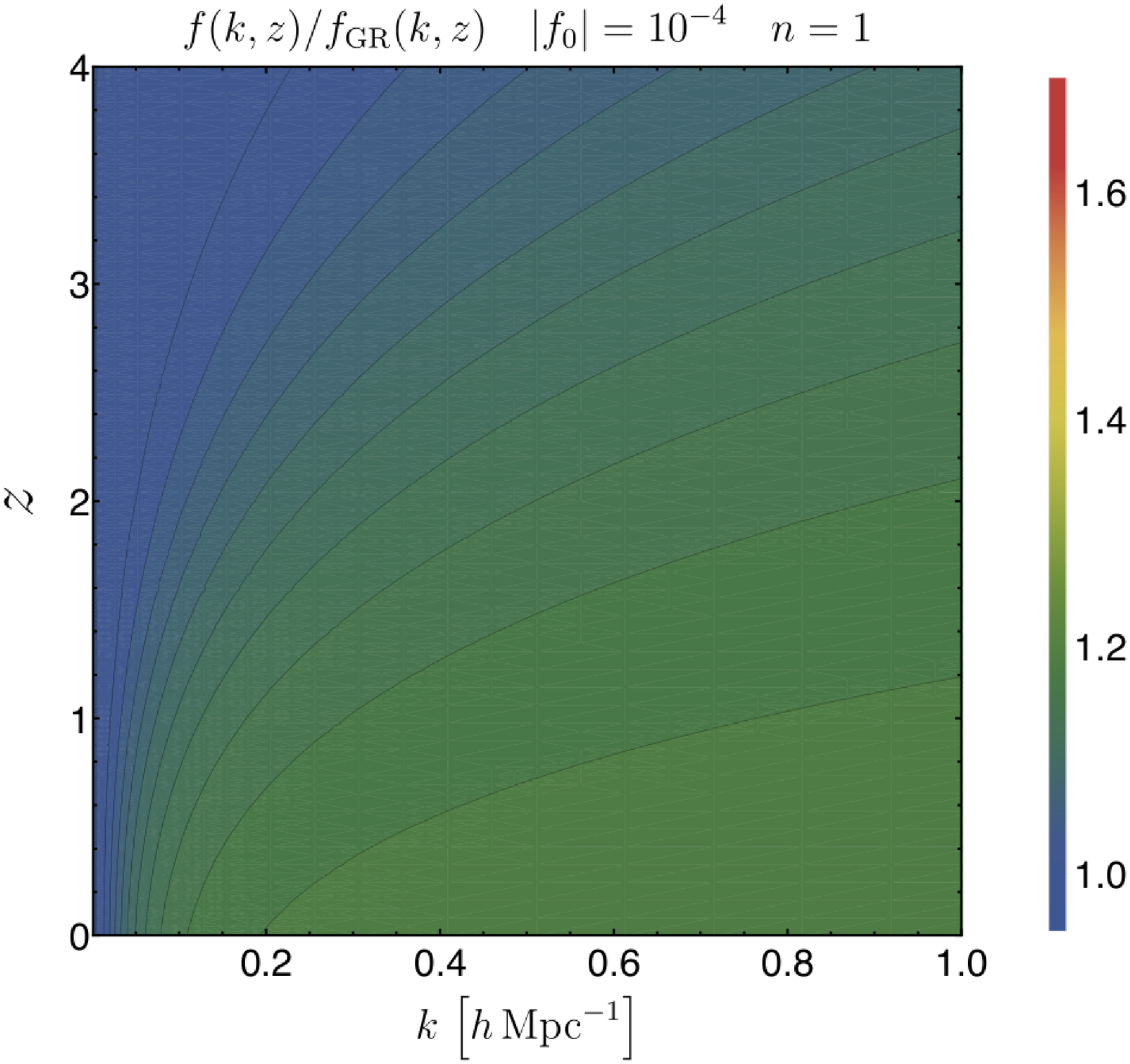}}} &
{\epsfxsize=5.25 cm \epsfysize=5.25 cm{\epsfbox[21 21 736 655]{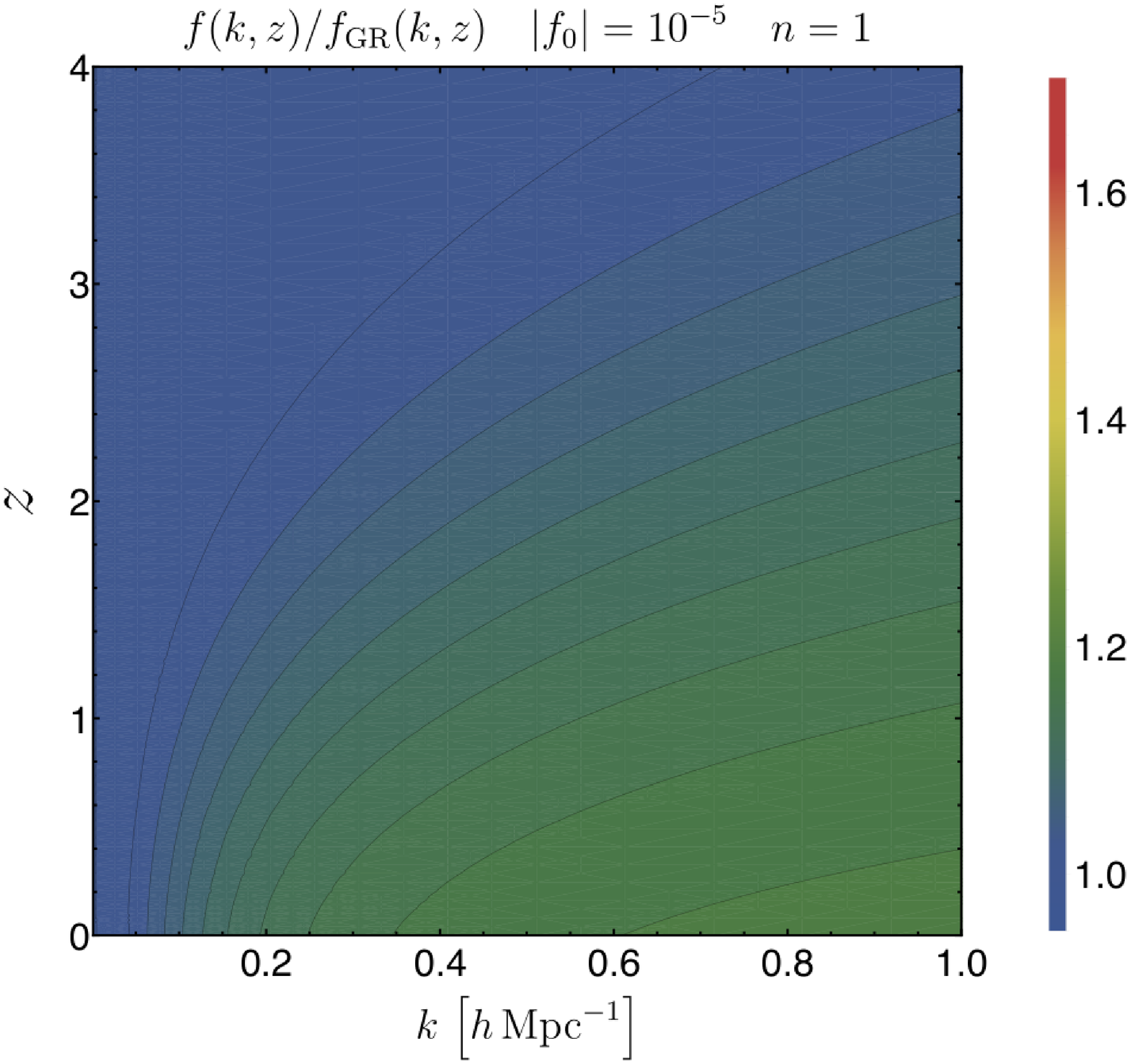}}} &
{\epsfxsize=5.25 cm \epsfysize=5.25 cm{\epsfbox[21 21 736 655]{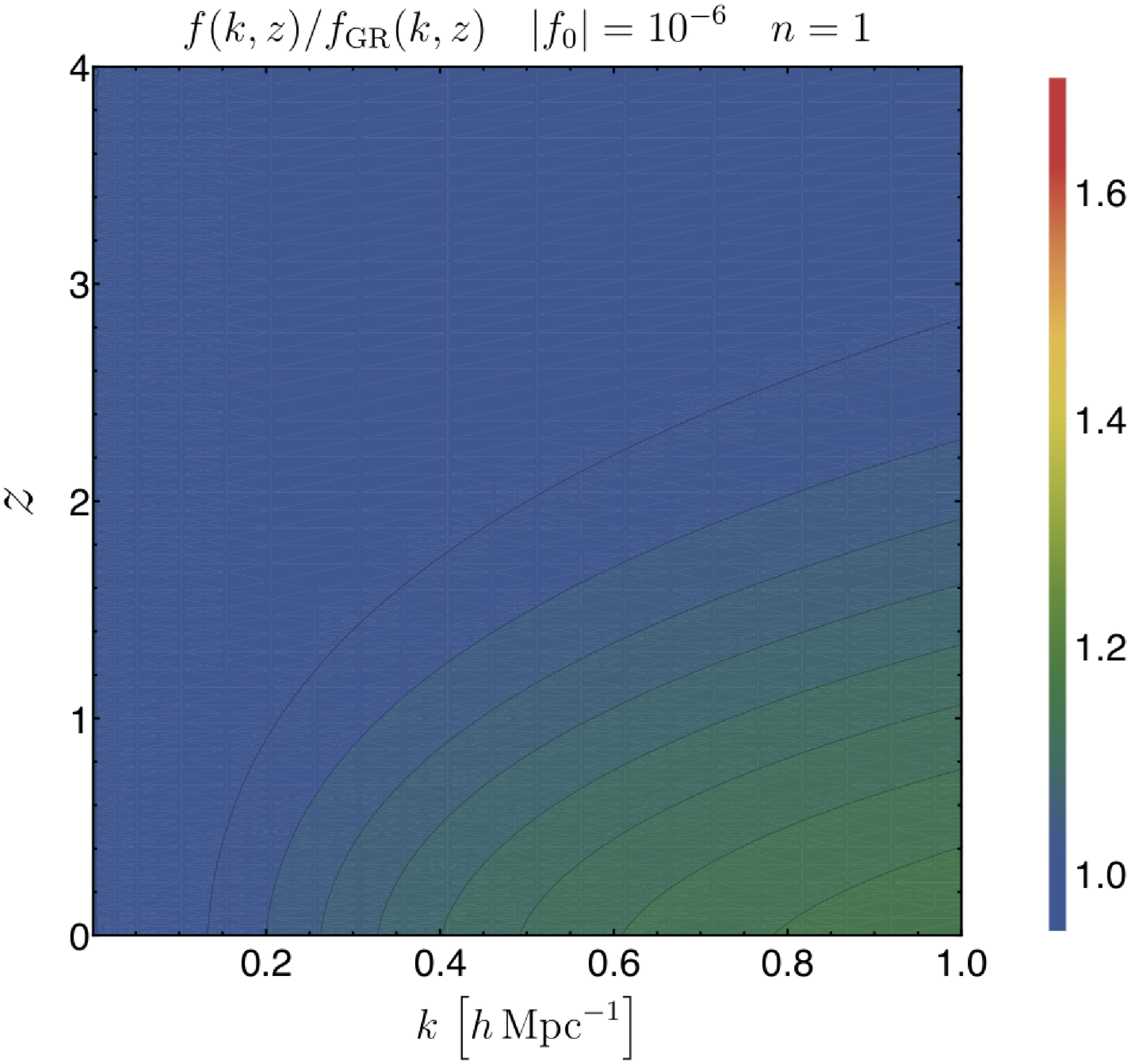}}} 
\end{tabular}
\caption{We plot the linear growth rate $f(k,z)=\partial\ln D_{+}(k,z)/ \partial\ln a$, 
normalised to the $\Lambda$-CDM value, for $f({\rm R})$ models as a function of the 
wave number $k$ (in units of $h \rm Mpc^{-1}$) and of the redshift $z$.
The functional form for the $f(R)$ model is described in Eq.(\ref{eq:fofR}), with $n=1$.
The parameter $f_{R_0}$ is fixed at $-10^{-4}$ (left-panel), $-10^{-5}$ (middle panel) 
and $-10^{-6}$ (right-panel), respectively.}
\label{fig:fkz_fofR}
\end{figure}

\begin{figure}
\centering
\vspace{0.15cm}
\begin{tabular}{cc}
{\epsfxsize=5.5cm \epsfysize=5.5cm{\epsfbox[21 21 736 655]{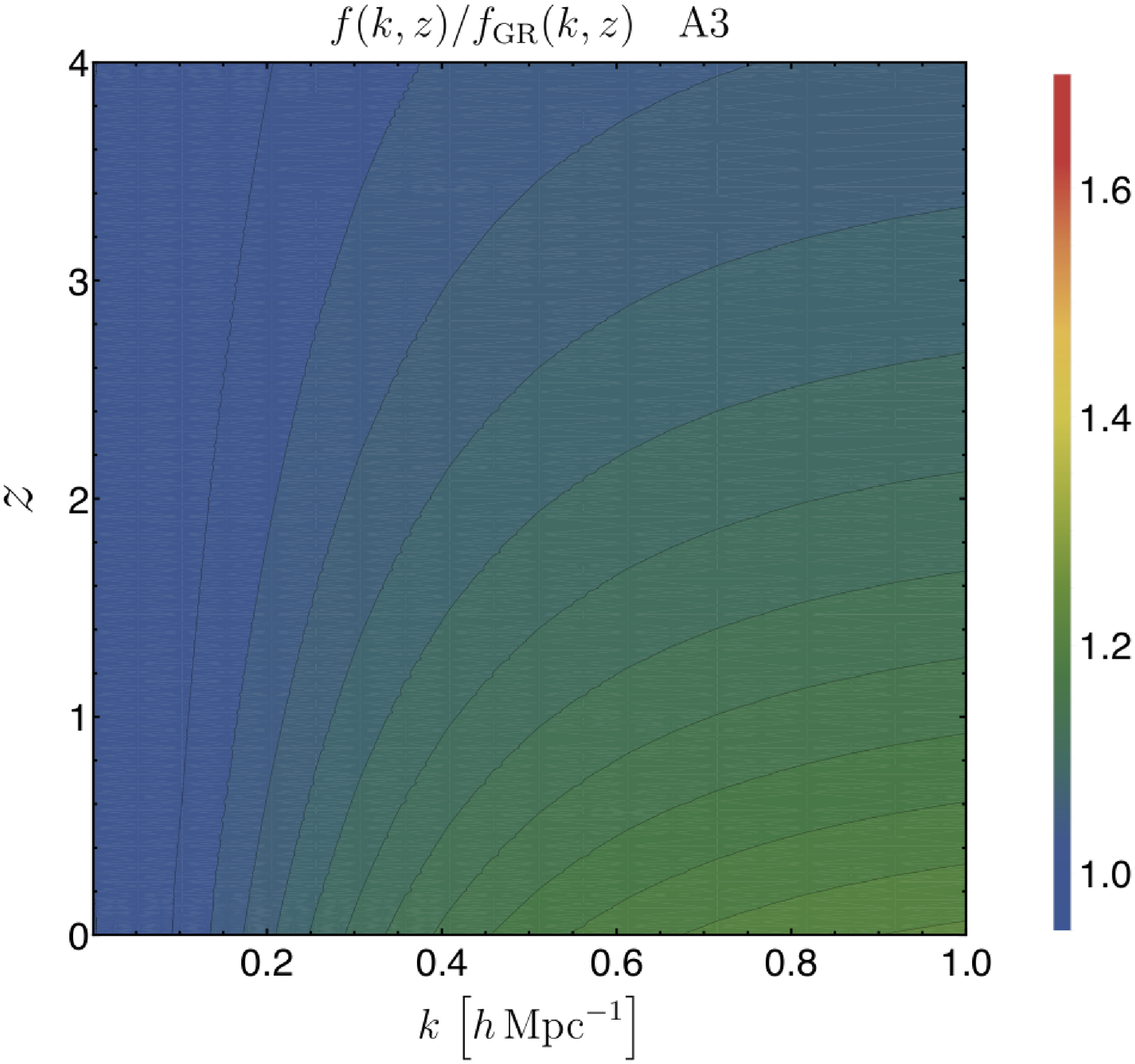}}} &
{\epsfxsize=5.5cm \epsfysize=5.5cm{\epsfbox[21 21 736 655]{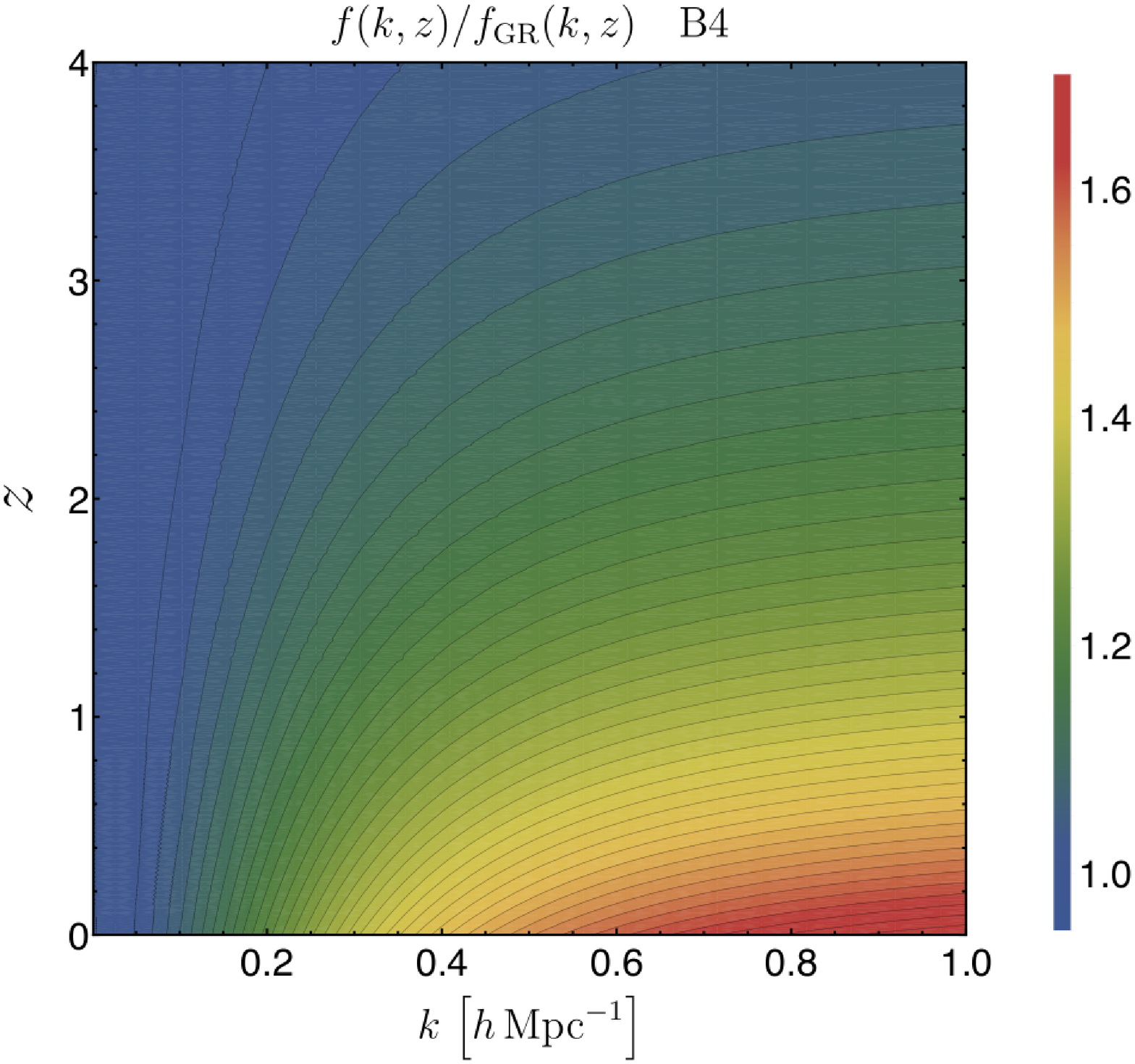}}} \\
{\epsfxsize=5.5cm \epsfysize=5.5cm{\epsfbox[21 21 736 655]{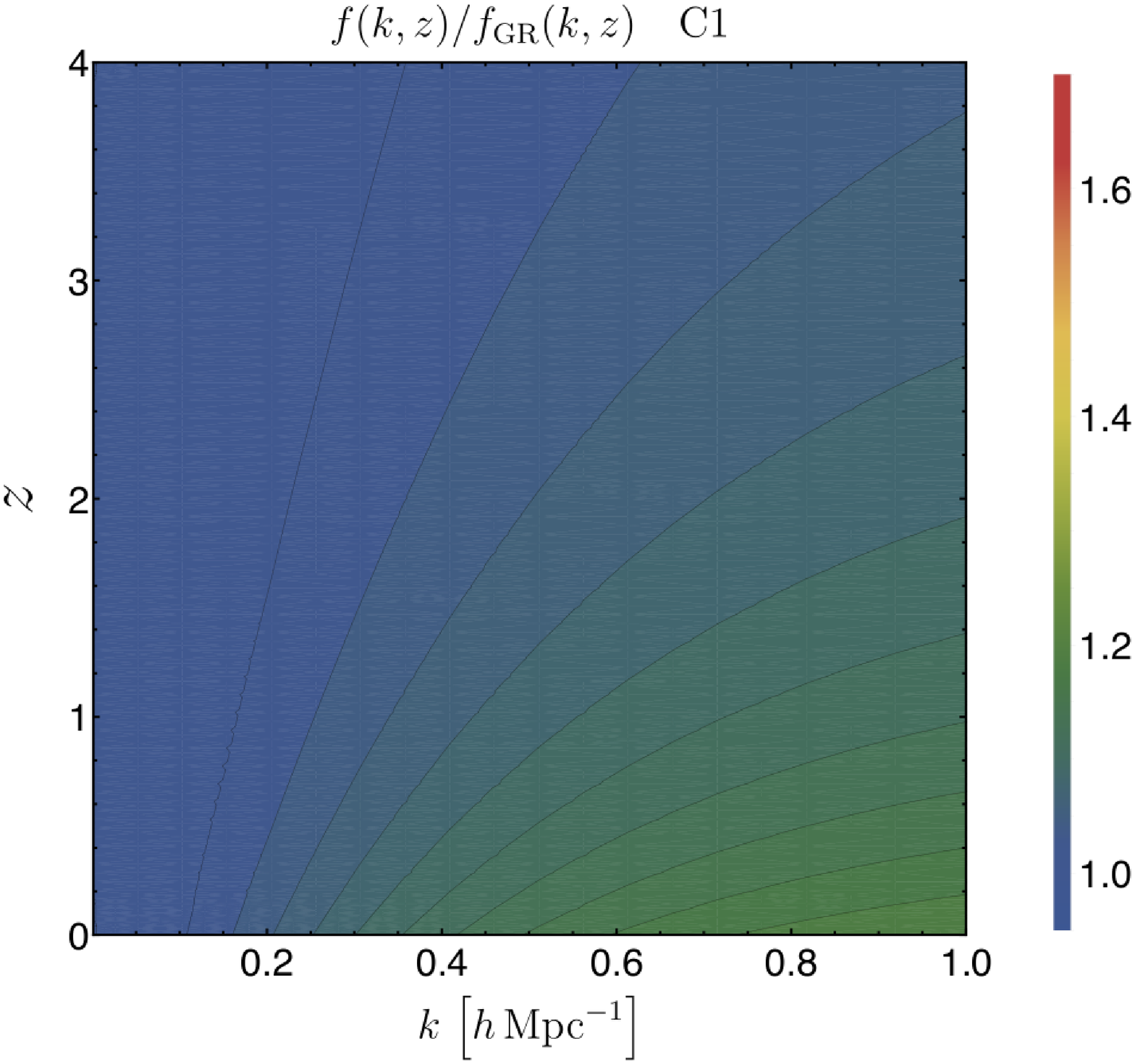}}} &
{\epsfxsize=5.5cm \epsfysize=5.5cm{\epsfbox[21 21 736 655]{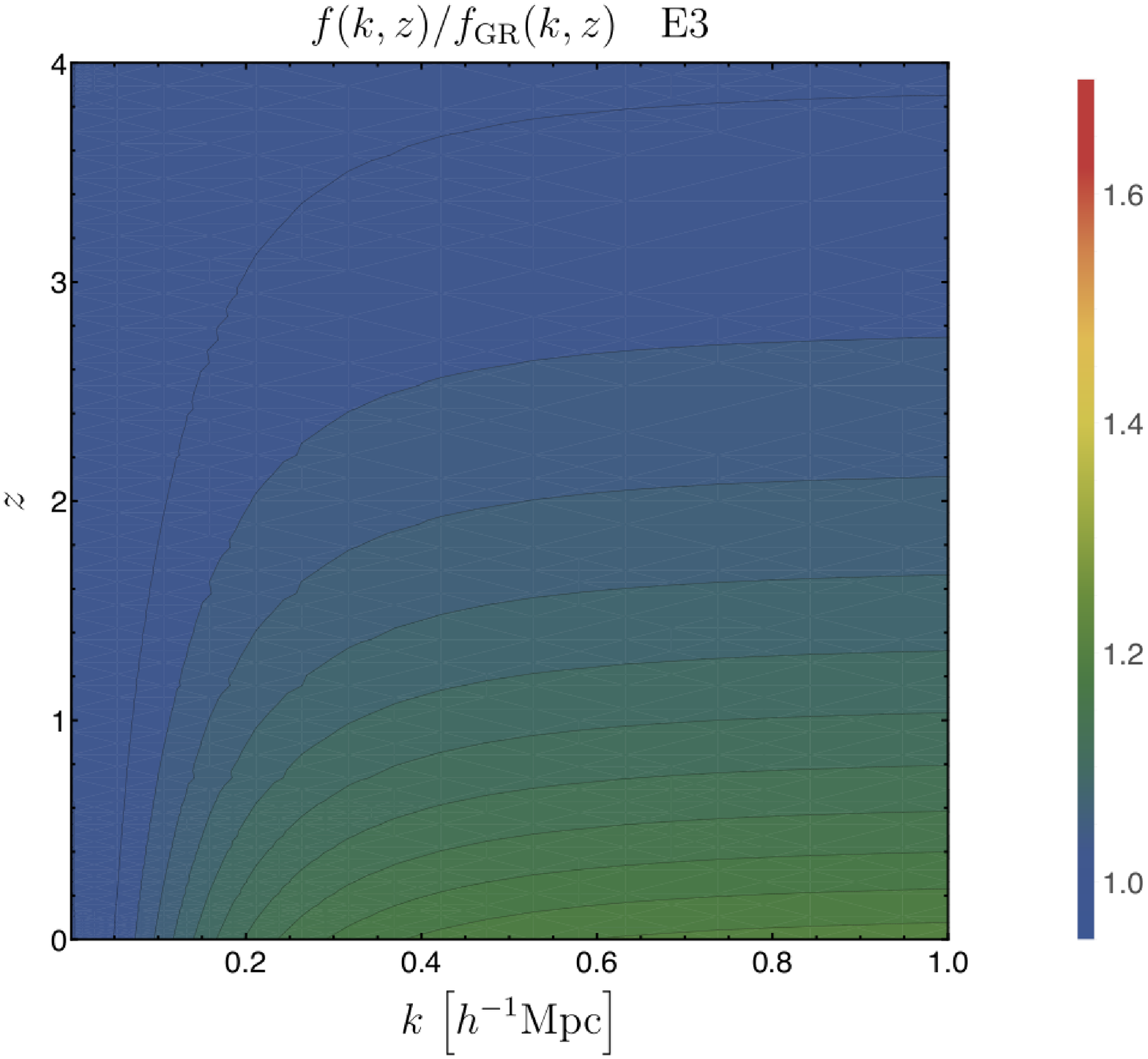}}} 
\end{tabular}
\caption{We plot the linear growth rate $f(k,z)=\partial \ln D_{+}(k,z)/ \partial \ln a$, 
normalised to the $\Lambda$-CDM value, as in Figure-\ref{fig:fkz_fofR} but for
the Dilaton models.
The top panels correspond to the Dilaton models A3 (left) and B4 (right) models.
The bottom panels correspond to the Dilaton models C1 (left) and E3 (right) 
respectively (i.e. only the extreme case from each class of models is shown).}
\label{fig:fkz_dilaton}
\end{figure}

In the modified gravity scenarios, the linear growing mode $D_+(k,t)$ and the
linear growth rate $f(k,t) = \partial \ln D_+ / \partial \ln a$ become scale dependent.
We show the linear growth rates $f(k,z)$ in Figure-\ref{fig:fkz_fofR} for the $f(R)$, 
$(n=1)$, models and in Figure-\ref{fig:fkz_dilaton} for the Dilaton models.
For the Dilaton models, we only show those models that exhibit maximal departure
from GR predictions.
The modified gravity models that we consider in this paper amplify and accelerate
the growth of large-scale structures, so that their linear growth rates are greater than
the $\Lambda$CDM one. The relative importance of the fifth force also increases with 
time and is mostly relevant in the dark energy era, at $z \lesssim 2$.
In addition, we can see in Figures-\ref{fig:fkz_fofR} and -\ref{fig:fkz_dilaton}
the scale dependence generated by these modified gravity models.
The deviation from the $\Lambda$CDM growth rates decreases along with 
$|f_{R_0}|$. For the dilaton models the deviation decreases on these scales
as $\rm{B4>A3>E3>C1}$.

\begin{figure}
\centering
\vspace{0.15cm}
{\epsfxsize=10.2 cm \epsfysize=5.2 cm{\epsfbox[38 355 550 594]{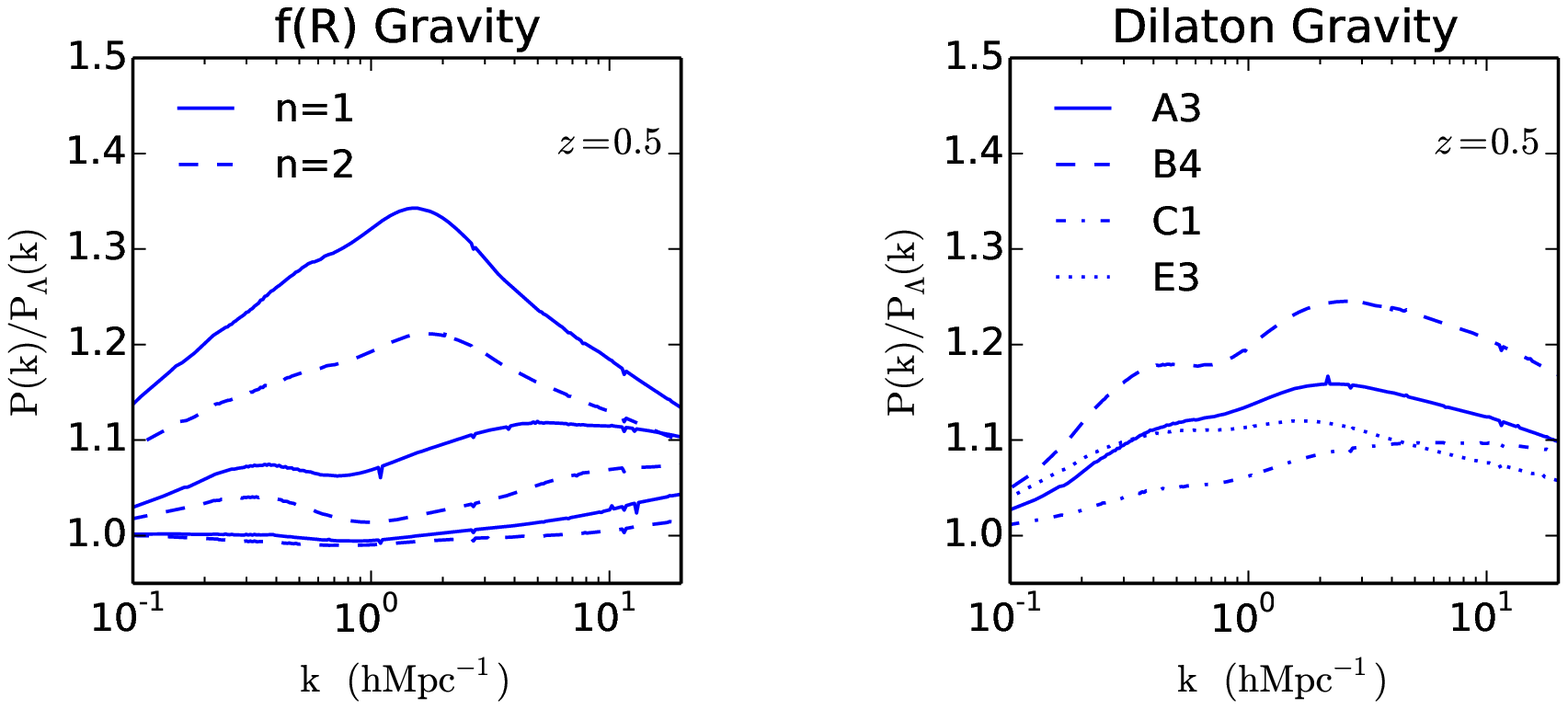}}}
\caption{The ratio of the matter power spectra in modified gravity theories and in the $
\Lambda$CDM model is plotted as a function of the wave number $k$. The left panel 
shows the results for $f(R)$ gravity and the right panel for Dilaton gravity models.
The redshift in both panels is fixed at $z=0.5$. We show two different models of $f(R)$ 
gravity that correspond to $n=1$ (solid-lines) and $n=2$ (dashed-lines). 
For each model we consider the three different amplitudes 
$f_{R_0}=-10^{-4}, -10^{-5}$ and $-10^{-6}$ (from top to bottom), respectively.
The parameters of the Dilaton models displayed in the right panel are given in 
Table-\ref{tabular:tab2}. Four different models are shown. Only the extreme version
(i.e., with the greatest deviation from $\Lambda$CDM) of each type of model is shown.}
\label{fig:ps_MG}
\end{figure}  

The non-linear matter power spectra for the modified gravity theories are displayed 
in Figure-\ref{fig:ps_MG}. 
The deviations from the $\Lambda$CDM power spectrum peak at the weakly non-linear
scales $k \sim 5 h \rm Mpc^{-1}$, due to the amplification by the non-linear dynamics.
Moreover, at very large scales (much beyond the Compton wave length)
all models converge back to GR and to the $\Lambda$CDM cosmology.
At high $k$, $k \gtrsim 10 h \rm Mpc^{-1}$, we may underestimate the deviation from
the $\Lambda$CDM power spectrum because we neglect the impact of the modified
gravity on the halo profiles. However, the analysis presented in this paper is restricted
to linear or weakly non-linear scales, $k \lesssim 0.2 h \rm Mpc^{-1}$, where the detailed
shape of halo profiles plays no role \citep{Val13} and our modelling of the matter density 
power spectrum is reliable.
In agreement with Figure-\ref{fig:fkz_fofR}, the deviations from the $\Lambda$CDM power
spectrum increase with $|f_{R_0}|$ and in the order $\rm{B4>A3>E3>C1}$.
For the $f(R)$ theories the deviations are smaller for a greater exponent $n$,
because this gives a faster increase of the scalar field mass $m$ at higher redshift,
as seen in Eq.(\ref{m-beta-fR}), whence a faster convergence to $\Lambda$CDM.

\section{Linear Redshift Space Distortions (beyond Kaiser Effect) in Modified Gravity}
\label{sec:red}

\subsection{Spherical Fourier-Bessel Formalism} 
\label{sec:sFB-real}

\subsubsection{Definition of the spherical Fourier-Bessel transform}
\label{sec:def-sFB}

Spherical coordinates are often a natural choice in the analysis of cosmological data sets as they can, by an appropriate choice of coordinates, be used to place the observer at the origin of the analysis. As upcoming surveys promise to yield both large (i.e. wide angle) and deep (i.e. large radial coverage) coverage of the sky, we require a simultaneous treatment of the extended radial coverage and spherical sky geometry. A natural basis for such an analysis is given by the spherical Fourier-Bessel formalism. In this section, we follow \citep{H03, CHK05, Rassat11, PM13} and detail the conventions used in this paper. 

Consider a homogeneous 3D random field $\Psi (r,\hat{\Omega})$ such that $\hat{\Omega}$ denotes the angular coordinate on the surface of a sphere and $r$ denotes the comoving radial distance. The eigenfunctions of the Laplacian will be constructed from products of the spherical Bessel functions of the first kind $j_{\ell} (kr)$ and spherical harmonics $Y_{\ell m} (\hat{\Omega})$ with eigenvalues of $-k^2$, for a 2-sphere. Assuming a flat background Universe, the sFB decomposition of the homogeneous 3D field will be given by
\ben
&& {\Psi} (r,\hat{\Omega}) = \sqrt{\frac{2}{\pi}} \int_0^{\infty}  dk \, \displaystyle\sum_{\lbrace \ell m \rbrace} \, \Psi_{\ell m} (k) \, k \, j_{\ell} (kr) \, Y_{\ell m} (\hat{\Omega}) , 
\een
\n
with the inverse relation given by
\ben
&& {\Psi}_{\ell m}(k) = \sqrt{\frac{2}{\pi}} \int_0^{\infty} dr \, r^2 \, \int d\hat{\Omega} \,\Psi(r, \hat{\Omega}) \, k \, j_\ell(kr) Y^{\ast}_{\ell m}(\hat{\Omega}).
\een
\n
This is something of a spherical polar analogue to the conventional Cartesian Fourier decomposition. 
In particular, defining the normalisation of the 3D Fourier transform and power spectrum as
\ben
&& \Psi({\bf r}) = \frac{1}{(2\pi)^{3/2}} \int d{\bf k} \, e^{{\rm ii} {\bf k} \cdot {\bf r}} \, \Psi({\bf k}) ,
\;\;\;
\langle \Psi({\bf k}) \Psi^{\ast}({\bf k}') \rangle = P(k) \delta_D({\bf k} - {\bf k}') ,
\een
the sFB coefficients and the Fourier modes can be related as
\ben
&& \Psi_{\ell m}(k) = {i}^{\ell} k \int d \hat{\Omega} \, Y_{\ell m}^{\ast}(\hat{\Omega}) \, \Psi(k,\hat{\Omega}) ,
\;\;\;
\Psi({k,\hat{\Omega}}) = \frac{1}{k} \displaystyle\sum_{\lbrace \ell m \rbrace} (-i)^{\ell} \Psi_{\ell m}(k) 
Y_{\ell m}(\hat{\Omega}) ,
\een
while the two power spectra obey
\ben
&& \langle \Psi_{\ell m}(k) \Psi^{\ast}_{\ell^{\prime}m^{\prime}}(k^{\prime}) \rangle = {\cal C}_{\ell} (k) \, 
\delta_D(k - k^{\prime}) \, \delta_{\ell \ell^{\prime}} \, \delta_{m m^{\prime}} 
\;\;\; \mbox{with} \;\;\; {\cal C}_{\ell}(k) = P(k) .
\een
\n
\subsubsection{Finite-depth surveys}
\label{sec:finite-depth-r}
In reality, we will often want to consider a cosmological random field that is only partially observed due to a finite survey volume. In this instance, we can construct the observed field $\tilde{\Psi} (r,\hat{\Omega})$,
which we denote with a tilde, by multiplying the original field $\Psi (r,\hat{\Omega})$ with a selection function $\varphi (r)$
\ben
&& \tilde{\Psi} (r,\hat{\Omega}) = \Psi (r,\hat{\Omega}) \, \varphi (r) .
\een
\n
The sFB coefficients of this finite-depth field can be related to those of the field $\Psi$ by
\ben
&& \tilde{\Psi}_{\ell m}(k) = \int_0^{\infty} d k' \, W_{\ell}(k,k') \Psi_{\ell m}(k')  \;\;\;
\mbox{with} \;\;\; W_{\ell}(k,k') = \frac{2}{\pi} \int_0^{\infty} dr \, r^2 \varphi(r) k j_{\ell}(kr) k' j_{\ell}(k'r) .
\label{Psi-tilde-W}
\een

The introduction of the selection function means that the homogeneity criterion is not valid in the radial direction and the observed sFB power spectrum will now given by
\ben
&& \langle \tilde{\Psi}_{\ell m} (k) \, \tilde{\Psi}^{\ast}_{\ell^{\prime} m^{\prime}} (k^{\prime}) \rangle = \tilde{\cal C}_{\ell} (k , k^{\prime}) \, \delta_{\ell \ell^{\prime}} \, \delta_{m m^{\prime}}  , \;\;\;
\mbox{with} \;\;\; \tilde{\cal C}_{\ell}(k,k') = \int_0^{\infty} dk'' \, W_{\ell}(k,k'') W_{\ell}(k',k'') \, P(k'') .
\een
\n
Typically we will often be interested in the diagonal modes for which $k = k^{\prime}$ as the sFB power spectrum falls off rapidly away from the diagonal. 
\subsection{Spherical Fourier-Bessel Formalism: Applications} 
As previously mentioned, the 3D approach will be particularly important for the analysis of cosmological data from future wide-field surveys. This is especially true for large angular scales in which the plane parallel approximation, the distant observer approximation or, equivalently, the high $\ell$ approximations to the spherical harmonics are all inadequate. Conventionally, redshift space distortions (RSD) are typically studied in 3D \citep{Fish95} whereas weak-lensing has traditionally been studied with projected surveys (2D) due to a lack of, or uncertain, photometric redshift information about individual sources. Due to the information available, many of these studies were also limited to small patches of the sky and therefore invoked the flat-sky approximation. With the advent of surveys that can provide accurate photometric redshift information across wide areas of the sky, there has been a growing demand for techniques that can make use of this full-sky 3D information. For example, an early approach to incorporating this 3D information into the 2D projected surveys was to invoke some form of tomographic reconstruction in which the sources are divided up into slices at different redshifts and a 2D analysis is performed in each of these redshift bins. More recently, \cite{H03} proposed a genuine 3D formalism for weak-lensing surveys based on the spherical Fourier-Bessel expansions. These studies were later extended to a detailed description of weak lensing observables on the full 3D sky \cite{CHK05}. In addition, power-spectrum estimation techniques have been generalised to the analysis of higher-order statistics in 3D \citep{MHC11,MKHC11}.

Another growing area of research within the 3D approach is the cross-correlation of galaxy and weak lensing surveys with other cosmological observables. For example, \cite{Shapiro11} studied the cross-correlation of 3D galaxy surveys with the projected CMB in order to study the integrated Sachs-Wolfe (ISW) effect. Likewise, \cite{PM14} detailed the cross-correlation of 3D weak-lensing with the projected thermal Sunyaev-Zel'dovich (tSZ) effect as a way to recover redshift information that is lost in the line-of-sight projection of the thermal pressure of free elections.

The 3D approach can also be used to study and characterise baryon acoustic oscillation (BAO) features in the matter power spectrum. The sFB approach was first used in this context by \cite{Rassat11} and later extended to include linear RSD as well as non-linear effects in the matter power spectrum by \cite{PM13}.
\subsection{Redshift Space Distortions}
\label{sec:RSD}
\subsubsection{Expansion to first order over peculiar velocities}
\label{sec:velocity}
The existence of inhomogeneous structure in the Universe induces peculiar velocities that 
lead to distortions in the observed clustering of galaxies as measured in redshift space. 
The anisotropies generated by these distortions are known as \textit{redshift space distortions} (RSD) 
which, together with bias and non-linear evolution, induce departures in the measured matter power spectrum 
away from the configuration-space power spectrum predicted by linear perturbation theory \citep{Kaiser87}. 
These distortions necessarily complicate the cosmological interpretation of spectroscopic galaxy surveys 
but the RSD are also one of the most promising probes for the measurement of the growth rate of structure 
formation and hence a useful probe for models of dark energy and modified theories of gravity.

The effect of the RSD on the matter power spectrum and clustering statistics can be broadly split into two effects: 
the linear Kaiser effect and the finger of God (FoG) effect. The linear Kaiser effect is a coherent distortion of 
the peculiar velocity along our line of sight with an amplitude controlled by the growth rate. The Kaiser effect 
leads to an enhancement of the power spectrum amplitude at small $k$ \citep{Kaiser87}. The FoG effect arises due 
to the random distribution of peculiar velocities for galaxies within virialized structures. These peculiar velocities 
lead to an incoherent contribution in which we have dephasing and a suppression of the clustering amplitude at high $k$ \citep{Jack72}. 
The effect of a peculiar velocity, or departure from the Hubble flow, ${\bf v}({\bf{r}})$ at ${\bf{r}}$ is to distort 
the observed comoving position in redshift space ${\bf{s}}$ from its true comoving position in real space ${\bf{r}}$:
\ben
&& {\bf{s}}({\bf{r}}) = {\bf{r}} + \frac{{\bf{v}}({\bf{r}}) \cdot \hat{\Omega}}{a H} .
\een
\n
In the following, we denote by a superscript ``s'' fields that are defined in redshift space ${\bf s}$,
to distinguish them from the real-space fields.
The redshift-space sFB transform can still be defined as in the real space case,
\ben
&& {\Psi}^s_{\ell m}(k) = \sqrt{\frac{2}{\pi}} \int_0^{\infty} ds \, s^2 \, \int d\hat{\Omega} \,\Psi^s(s, \hat{\Omega}) \, k \, j_\ell(ks) Y^{\ast}_{\ell m}(\hat{\Omega}).
\een
The conservation of matter implies $[1 + \delta^s ({\bf{s}}) ] d{\bf{s}} = [ 1 + \delta ({\bf{r}}) ] d{\bf{r}}$,
so that in the case of the sFB transform of the density contrast we can make the change of
integration variable from ${\bf s}$ to ${\bf r}$, as
\ben
&& \delta^s_{\ell m}(k) = \sqrt{\frac{2}{\pi}} \int_0^{\infty} dr \, r^2 \, \int d\hat{\Omega} \,[1+\delta(r,\hat{\Omega})] \, k \, j_\ell(ks) Y^{\ast}_{\ell m}(\hat{\Omega}) - \sqrt{\frac{2}{\pi}} \int_0^{\infty} ds \, s^2 \, \int d\hat{\Omega} \, k \, j_\ell(ks) Y^{\ast}_{\ell m}(\hat{\Omega}) .
\een
The last term only contributes for the monopole ($\ell=m=0$). 
Expanding over the peculiar velocity, we obtain
\ben
&& \{\ell,m\} \neq \{0,0\} : \;\;\;
\delta^s_{\ell m}(k) = \sqrt{\frac{2}{\pi}} \int_0^{\infty} dr \, r^2 \, \int d\hat{\Omega} \,[1+\delta(r,\hat{\Omega})] \left[ k \, j_\ell(kr) + \frac{{\bf{v}}({\bf{r}}) \cdot \hat{\Omega}}{a H} k^2 j_{\ell}'(kr) + ... \right] 
Y^{\ast}_{\ell m}(\hat{\Omega}) 
\label{delta-s-lm-neq-00}
\een

Using the following perturbative expansion over powers of the peculiar velocity:
\ben
&& \delta^s_{\ell m}(k) = \delta^{s(0)}_{\ell m}(k) + \delta^{s(1)}_{\ell m}(k) + ... ,
\label{delta-s-expand-0-1}
\een
we have
\ben
&& \{\ell,m\} \neq \{0,0\} : \;\;\; \delta^{s(0)}_{\ell m}(k) = \delta_{\ell m}(k) =
\sqrt{\frac{2}{\pi}} \int_0^{\infty} dr \, r^2 \, \int d\hat{\Omega} \, \delta(r,\hat{\Omega})  k \, j_\ell(kr) 
Y^{\ast}_{\ell m}(\hat{\Omega}) 
= i^{\ell} k \int d\hat{\Omega} \, \delta(k,\hat{\Omega})  Y_{\ell m}^{\ast}(\hat{\Omega}) 
\label{delta-s0-def}
\een
and
\ben
&& \delta^{s(1)}_{\ell m}(k) = \sqrt{\frac{2}{\pi}} \int_0^{\infty} dr \, r^2 \, \int d\hat{\Omega} \, 
[1+\delta(r,\hat{\Omega})] \frac{{\bf{v}}({\bf{r}}) \cdot \hat{\Omega}}{a H} k^2 j_{\ell}'(kr)
Y^{\ast}_{\ell m}(\hat{\Omega}) 
\een
The continuity equation reads as $a \partial \delta/\partial t + \nabla [ (1+\delta) {\bf v} ]=0$.
Even though $(1+\delta) {\bf v}$ is not curl-free, to obtain an order of magnitude estimate
we may write
\ben
&& (1+\delta) {\bf v} \simeq - a \nabla^{-1} \frac{\partial \delta}{\partial t} ,
\label{divergence-inverse}
\een
which is exact at linear order, and
\ben
\delta^{s(1)}_{\ell m}(k) & \simeq & - \sqrt{\frac{2}{\pi}} \int_0^{\infty} dr \, r^2 \, \int d\hat{\Omega} \, 
\left( \nabla^{-1} \frac{\partial \delta}{\partial t} \right) (r,\hat{\Omega}) \cdot \frac{\hat{\Omega}}{H} 
k^2 j_{\ell}'(kr) Y^{\ast}_{\ell m}(\hat{\Omega}) \\
& \simeq & \frac{2}{\pi} i^{\ell} \frac{k^2}{H} \int_0^{\infty} dk' \, k' \int d\hat{\Omega} 
\frac{\partial\delta(k',\hat{\Omega})}{\partial t} \int_0^{\infty} dr \, r^2 j'_{\ell}(kr) j'_{\ell}(k'r) Y_{\ell m}^{\ast}(\hat{\Omega}) . 
\label{delta-s-1-dt}
\een
In the linear regime, the growth rate of the linear growing mode $D_+(k,t)$
(which usually depends on the wave number in modified-gravity scenarios) is defined as
\ben
&& f(k,t) = \frac{\partial\ln D_+}{\partial\ln a} = \frac{1}{2} \frac{\partial\ln P_L}{\partial\ln a} , \;\;\;
\mbox{and} \;\;\; \frac{\partial\delta_L({\bf k},t)}{\partial t} = f H \delta_L .
\een
In a similar fashion, we define the nonlinear growing mode $D_P(k,t)$ and growth rate $f_{P}(k,t)$ from the non-linear power spectrum
as
\ben
&& D_P(k,t) = \sqrt{ \frac{P(k,z)}{P_L(k,0)} } , \;\;\; 
f_{P}(k,t) =  \frac{1}{2} \frac{\partial\ln P}{\partial\ln a} ,
\label{DP-fP-def}
\een
where $P_L$ and $P$ are the linear and non-linear power spectra,
and we use in Eq.(\ref{delta-s-1-dt}) the approximation
\ben
&& {\partial\delta({\bf k},t) \over \partial t} \simeq f_{P} H \delta ,
\label{fP-dt}
\een
which is exact at linear order. This gives
\ben
&& \delta^{s(1)}_{\ell m}(k) \simeq \frac{2}{\pi} i^{\ell} k^2  \int_0^{\infty} dk' k' \int d\hat{\Omega}\, Y_{\ell m}^{\ast}(\hat{\Omega})\,
\int_0^{\infty} dr \, r^2 \delta(k',\hat{\Omega})  f_P(k',z) j'_{\ell}(kr) j'_{\ell}(k'r) . 
\label{delta-s1-Fourier}
\een
Thus, Eqs.(\ref{delta-s0-def}) and (\ref{delta-s1-Fourier}) give the exact expression of the
matter density sFB transform in redshift space at linear order.
Moreover, it includes a partial account of nonlinear contributions. The zeroth-order term
Eq.(\ref{delta-s0-def}) over velocities includes all nonlinear contributions from gravitational clustering.
The first-order term Eq.(\ref{delta-s1-Fourier}) over velocities is exact at linear order but only approximate
 at nonlinear order, because of the approximations Eq.(\ref{divergence-inverse}) and Eq.(\ref{fP-dt}).
Nevertheless, they should capture the magnitude of nonlinear contributions to this large-scale
 Kaiser effect.
 
Although we go beyond the usual linear-order approximation, by taking into account this nonlinear
 contributions, we restrict ourselves to large weakly nonlinear scales, because we neglected
 small-scale virial motions associated with the fingers-of-god effect. 
 Thus, the redshift-space distortions considered in this paper correspond to the Kaiser effect
 associated with large-scale coherent flows.
\subsubsection{Finite-depth surveys}
\label{sec:finite-depth-s}
In the case of finite-depth surveys, as in \textsection\ref{sec:finite-depth-r}, we need to multiply the density
field by the selection function $\varphi(r)$. We still define the selection function in real space
rather than redshift space because it is not necessarily affected by peculiar velocities in the same
fashion as the radial coordinate. Typically, the selection function depends on the flux and observed angular
size of the objects and writing it as a function of redshift is a convenient approximation. Besides,
it typically varies on cosmological scales, of order $c/H$, which are much greater than the weakly 
nonlinear scales that we aim to probe (or order $2\pi/k$). Then, one can write $\varphi(r) \simeq \varphi(s)$,
which also corresponds to neglecting logarithmic radial gradients of the selection function as compared with
logarithmic gradients of the density field.

Then, Eq.(\ref{delta-s-lm-neq-00}) becomes
\ben
&& \{\ell,m\} \neq \{0,0\} : \;\;\;
\tilde{\delta}^s_{\ell m}(k) = \sqrt{\frac{2}{\pi}} \int_0^{\infty} dr \, r^2 \, \int d\hat{\Omega} \,[1+\delta(r,\hat{\Omega})] 
\, \varphi(r) \, \left[ k \, j_\ell(kr) + \frac{{\bf{v}}({\bf{r}}) \cdot \hat{\Omega}}{a H} k^2 j_{\ell}'(kr) + ... \right] 
Y^{\ast}_{\ell m}(\hat{\Omega}) 
\label{delta-s-lm-neq-00-phi}
\een
We can again expand over the peculiar velocity as in Eq.(\ref{delta-s-expand-0-1}).
The zeroth-order component is given by the real-space expression (\ref{Psi-tilde-W}),
\ben
&& \tilde{\delta}^{s(0)}_{\ell m}(k) =  \int_0^{\infty} d k' \, W^{s(0)}_{\ell}(k,k') \delta_{L\ell m}(k',0)  \;\;\;
\mbox{with} \;\;\; \\
&& W^{s(0)}_{\ell}(k,k') 
= \frac{2}{\pi} \int_0^{\infty} dr \, r^2 D_P(k',z) \varphi(r)\, k \, j_{\ell}(kr)\, k'\, j_{\ell}(k'r) ,
\label{delta-tilde-s0-W}
\een
where we defined $\delta_{L\ell m}(k',0)$ the linear density contrast today, at $z=0$,
and $D_P(k,z)$ is the non-linear growing mode defined in 
Eq.(\ref{DP-fP-def}), while the first-order component reads as
\ben
&& \tilde{\delta}^{s(1)}_{\ell m}(k) = \int_0^{\infty} d k' \, W^{s(1)}_{\ell}(k,k') \delta_{L\ell m}(k',0)  \;\;\;
\mbox{with} \;\;\; W^{s(1)}_{\ell}(k,k') = \frac{2}{\pi}  k^2 \int_0^{\infty} dr \, r^2 D_P(k',z)
f_{P}(k',z) \,  \varphi(r) j'_{\ell}(kr) \, j'_{\ell}(k'r) .
\label{delta-tilde-s1-W}
\een
We again define the auto- and cross-power spectra constructed from these harmonic coefficients as
\ben
&& \langle \tilde{\delta}^{s(\alpha)}_{\ell m} (k) \tilde{\delta}^{s(\beta) \ast}_{\ell^{\prime}m^{\prime}} (k^{\prime}) \rangle = 
\tilde{\myC}^{s(\alpha ,\beta)}_\ell (k,k^{\prime}) \delta_{\ell \ell^{\prime}} \delta_{m m^{\prime}} ,
\een
where the indices $\alpha$ and  $\beta$ take values $\{ 0,1 \}$, associated with the first two orders
of the expansion over peculiar velocities. 
From Eqs.(\ref{delta-tilde-s0-W}) and (\ref{delta-tilde-s1-W}) we obtain
\ben
&& \tilde{\myC}^{s(\alpha , \beta)}_{\ell} (k,k') = \int d k'' \, W_{\ell}^{s(\alpha)}(k,k'') W_{\ell}^{s(\beta)}(k',k'')P_L(k'',0),
\label{eq:PowSpec}
\een
and, up to first order over peculiar velocities, the redshifted power spectrum reads as
\ben
&& \tilde{\myC}_{\ell}^s(k,k') = \tilde{\myC}^{s(00)}_{\ell} (k,k') + \tilde{\myC}^{s(10)}_{\ell} (k,k') 
+ \tilde{\myC}^{s(01)}_{\ell} (k,k') . 
\label{eq:cls_def}
\een
The covariance matrix $\tilde{\myC}^{s}_{\ell}(k,k')$ is symmetric and approximately block diagonal in nature.

The kernels $W_{\ell}^{s(\alpha)}$ are exact at linear order over the matter density fluctuations and contain some non-linear contributions.
Thus, the real-space power $\tilde{\myC}^{s(0,0)}_{\ell}$ is exact at full 
non-linear order, within the model that we use to compute the non-linear matter power
spectrum, but the cross-power $\tilde{\myC}^{s(0,0)}_{\ell}$ is only exact up to
linear order, because it involves the approximations Eq.(\ref{divergence-inverse}) and
Eq.(\ref{fP-dt}). 
To check that the non-linear contributions do not play a significant role, and do
not degrade our predictions, we compare our results with those associated with
the linear power spectrum itself.
Thus, we introduce the kernels ${W}^{s({0})}_{L\ell}$ and ${W}^{s({1})}_{L\ell}$ that source the 3D linear power spectrum ${P_{\rm L}(k^{\prime},0)}$
and depend on the linear growth rate $D_+$ respectively through the following convolutions:
\ben
&&{W}^{s({0})}_{L\ell} (k,k^{\prime}) = {2 \over \pi} 
\int_0^{\infty} dr \, r^2 \, \cb{D_{+}(k^{\prime} , z)}  \, \varphi(r) \, k \, j_\ell (kr) k^{\prime} \, j_\ell (k^{\prime} r);
 \label{eqn:Il2} \quad\\
&& {W}^{s({1})}_{L\ell} (k,k^{\prime}) = {2 \over \pi}
 k^2 \int dr \, r^2 \, \cb{D_{+}(k^{\prime},z)} \, 
f(k',z) \, \varphi(r) j^{\prime}_\ell (kr)\,  j^{\prime}_\ell (k^{\prime} r). 
\label{eqn:UU}
\een
These are the linear counterparts of the kernels ${W}^{s({0})}_{\ell}$ and 
${W}^{s({1})}_{\ell}$ defined in Eqs.(\ref{delta-tilde-s0-W}) and (\ref{delta-tilde-s1-W}).
The corresponding sFB linear power spectrum is defined through the following equation:
\ben
&& \tilde{\cal C}^{s(\alpha,\beta)}_{L\ell}(k,k') = \int d k'' \, W_{L\ell}^{s(\alpha)}(k,k'') 
W_{L\ell}^{s(\beta)}(k',k'')P_L(k'',0).
\label{eq:powUU}
\een
\subsubsection{Galaxy bias}
\label{sec:galaxy-bias}
In practice, we do not observe the matter density field itself, except in weak-lensing surveys,
but the galaxy distribution. Writing the galaxy density field as a linear function of the matter density
field with a scale-independent bias, $\delta_g({\bf r},z) = b(z) \delta({\bf r},z)$, while the velocity field remains
unbiased, we recover the same expressions as in the previous section, but with the kernels
$W_{\ell}^{gs(\alpha)}$ given by
\ben
&& W_{\ell}^{gs(0)}(k,k') = b(z) W_{\ell}^{s(0)}(k,k') , \;\;\;
W_{\ell}^{gs(1)}(k,k') = W_{\ell}^{s(1)}(k,k') ,
\label{Wg-def}
\een
where again the first-order term $W_{\ell}^{gs(1)}$ is only exact up to linear order
over the density and velocity fluctuations.
The same relations hold between the linear kernels $W_{L\ell}^{gs(\alpha)}$ 
and $W_{L\ell}^{s(\alpha)}$.
Then, the expansion (\ref{eq:cls_def}) is also a first order expansion over the ratio 
$\beta_P = f_{P}/b$. 
\section{\cb{Results}}
If we ignore the effects introduced by the selection function, i.e. set $\varphi(r)=1$, and 
neglect redshift space distortions, i.e. set $s=r$, then we recover the result of \cite{CHK05} 
for the un-redshifted contributions $\myC_{\ell} (k,k) = P_{\delta \delta} (k)$. 
These expressions hold for surveys with all-sky coverage. In the presence of homogeneity 
and isotropy, the 3D power spectrum will be independent of the angular multipole $\ell$. 
The introduction of a sky mask breaks isotropy and introduces additional mode-mode 
couplings. The machinery for dealing with partial sky coverage and a sky mask is reviewed 
in the appendix of \cite{PM13}. Note that in the above equations we neglect a number of 
additional non-linear terms. These include General Relativistic corrections, velocity terms 
as well as lensing contributions. In addition, the flat sky limits of Eq.(\ref{eq:PowSpec}) can 
be found in \citep{PM13}.  

For $\Lambda$CDM, $\beta(k,z) \approx \beta (z)$ but in modified theories of gravity 
$\beta(k,z) = f(k,z)/b(z)$ is typically $k$-dependant. For both $\Lambda$CDM and the 
modified theories of gravity considered in this paper, we use the full $\beta_P(k,z)$ taken 
from the numerical calculations of the non-linear matter power spectrum. 

We choose a survey with selection function $\varphi(r)\propto \exp(-r^2/r_0^2)$
and $r_0=150 h^{-1}\rm Mpc$. For the bias we use the model $b(z)=\sqrt{1+z}$. 
However, we find that our results are not very sensitive to $b(z)$ for the survey 
configuration that we have considered.
The results are presented for all-sky coverage.
\subsection{Impact of non-linear contributions and of redshift-space distortions}
\begin{figure}
\centering
{\epsfxsize=15.5 cm \epsfysize=4.75 cm{\epsfbox[30 425 550 590]{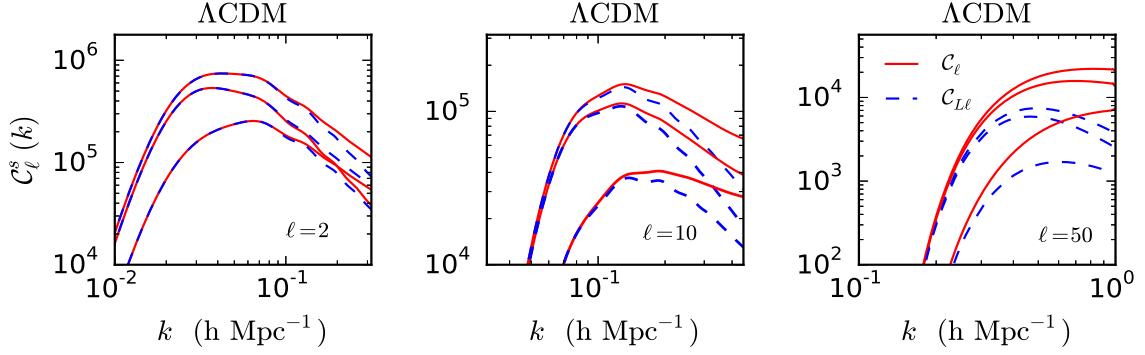}}} 
\caption{The 3D power spectra 
$\tilde{\cal C}^s_{\ell}(k)=\tilde{\cal C}^{s(00)}_{\ell}+\tilde{\cal C}^{s(10)}_{\ell}+\tilde{\cal C}^{s(01)}_{\ell}$ 
for $\Lambda$CDM are shown as a 
function of the radial wave number $k$ for various angular harmonics. 
From left to right the  panels represent $\ell=2, 10$ and $50$, respectively.
The topmost curve in each panel corresponds to $\tilde{\cal C}^s_{\ell}(k)$,
from Eq.(\ref{eq:cls_def}). Next, the two
sets of curves, from top to bottom, correspond to the contributions from 
$\tilde{\cal C}^{s(00)}_{\ell}$ and $\tilde{\cal C}^{s(10)}_{\ell}+\tilde{\cal C}^{s(01)}_{\ell}$,
respectively. In addition to the non-linear power $\tilde{\cal C}^s_{\ell}(k)$ (solid-lines) 
defined by the window function (\ref{eq:PowSpec}), we also
show the linear power $\tilde{\cal C}^s_{L\ell}(k)$ (lower-dashed lines) 
defined by the window function (\ref{eq:powUU}).}
\label{fig:lambda}
\end{figure} 
We show the diagonal entries of the 3D sFB power spectrum $\tilde{\cal C}^s_{\ell}(k,k)$,
as defined in Eq.(\ref{eq:cls_def}), in Figure-\ref{fig:lambda}, for the $\Lambda$CDM 
cosmology.
We also plot separately the contributions $\tilde{\cal C}^{s(00)}_{\ell}$
(equal to the real-space power) and 
$\tilde{\cal C}^{s(10)}_{\ell}+\tilde{\cal C}^{s(01)}_{\ell}$
(the redshift space contribution at first order over the peculiar velocity power spectrum).
In each case we also display the linear power spectra defined by Eq.(\ref{eq:powUU}).

We find that the non-linear contributions are quite small below
$k \lesssim 0.2 h$Mpc$^{-1}$ but cannot be neglected at higher $k$.
This transition scale also corresponds to the harmonic $\ell \gtrsim 10$.
For higher $\ell$ the non-linear contribution becomes more pronounced at lower $k$. 
The redshift space distortions typically give a $20\%$ contribution to the
total power on the scales displayed in Figure-\ref{fig:lambda}.
\subsection{Off-diagonal terms of the power spectrum matrix $\tilde{\cal C}^{s}_{\ell}(k,k')$}
\begin{figure}
\centering
\vspace{0.5cm}
\begin{tabular}{cccc}
{\epsfxsize=10.25 cm \epsfysize=4.75 cm{\epsfbox[31 399 388 590]{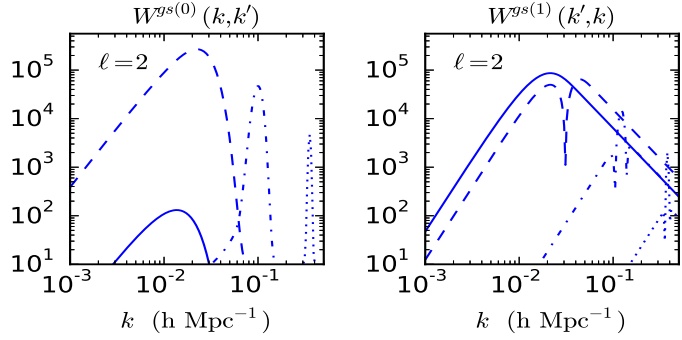}}} 
\end{tabular}
\caption{The kernel functions introduced in Eq.(\ref{Wg-def}) are presented. 
The left panel shows $W^{gs(0)}(k,k')$ and the right panel $W^{gs(1)}(k,k')$ for $\ell=2$. 
Different curves with peak position shifting from left to right correspond to 
$k^{\prime}=0.001\;(\rm solid\; lines), 0.025\; (\rm dashed\; lines),0.1\; (\rm dot-dashed\; lines),0.35 \; (dotted \;lines) \; h\, \rm{Mpc}^{-1}$.}
\label{fig:kernel}
\end{figure} 
\begin{figure}
\centering
\vspace{0.5cm}
\begin{tabular}{cccc}
{\epsfxsize=10.25 cm \epsfysize=4.75 cm{\epsfbox[31 399 388 590]{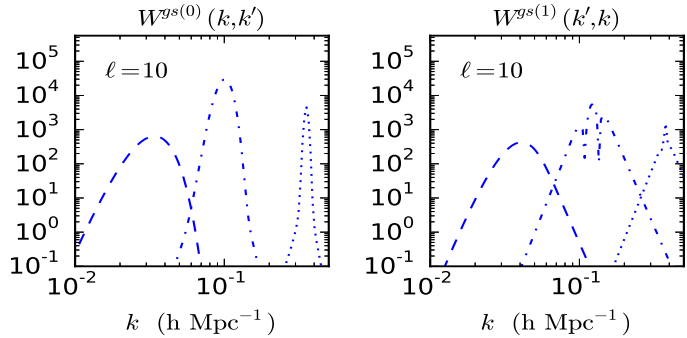}}} 
\end{tabular}
\caption{Same as Figure-\ref{fig:kernel} but for $\ell=10$.}
\label{fig:kernel10}
\end{figure} 
A few slices through the window functions $W_{\ell}^{gs(0)}(k,k')$ and 
$W_\ell^{gs(1)}(k,k')$ that represent mode-mixing in the presence of the radial 
selection function are displayed in Figure-\ref{fig:kernel} for $\ell=2$ and Figure-\ref{fig:kernel10} for $\ell=10$. 
For higher $\ell$ and $k'$ the mode mixing is more suppressed, that is, the window function is more 
strongly peaked around $k' \simeq k$.
\begin{figure}
\centering
\vspace{0.15cm}
\begin{tabular}{cccc}
{\epsfxsize=10.25 cm \epsfysize=5.2 cm{\epsfbox[31 399 388 600]{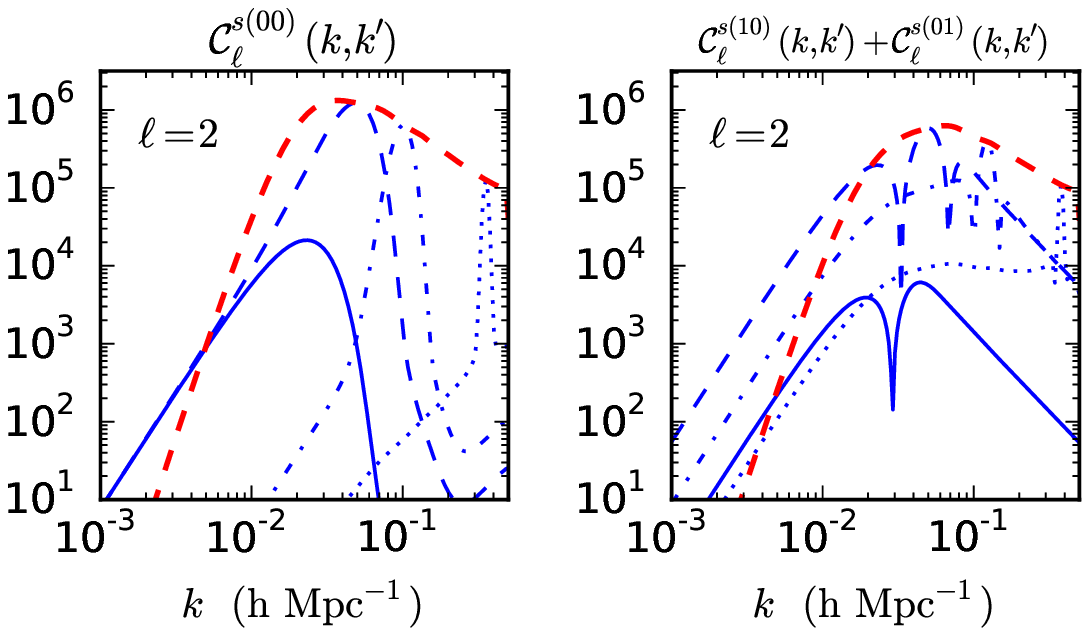}}} 
\end{tabular}
\caption{The slices of the covariance matrices are shown for $\ell=2$. The left panel 
shows $\tilde{\cal C}^{s(00)}_{\ell}(k,k')$ as a function of $k$ for fixed values of $k'$. 
The right panel corresponds to 
$\tilde{\cal C}^{s(10)}_{\ell}(k,k')+\tilde{\cal C}^{s(01)}_{\ell}(k,k')$. 
Various solid curves represent different fixed values of $k^{\prime}$, with 
$k^{\prime} = 0.005 (\rm solid\; lines), 0.05 (\rm dashed\; lines), 0.1 (\rm dot-dashed\; lines)$ and $0.35\; h\,{\rm Mpc}^{-1} (\rm dotted\;lines)$ from left to right. 
The thick dashed lines in each panel represent the diagonal entries of the covariance matrices,
$\tilde{\cal C}^{s}_{\ell}(k,k)$.} 
\label{fig:cov2}
\end{figure} 
\begin{figure}
\centering
\vspace{0.15cm}
\begin{tabular}{cccc}
{\epsfxsize=10.25 cm \epsfysize=5.2 cm{\epsfbox[31 399 388 600]{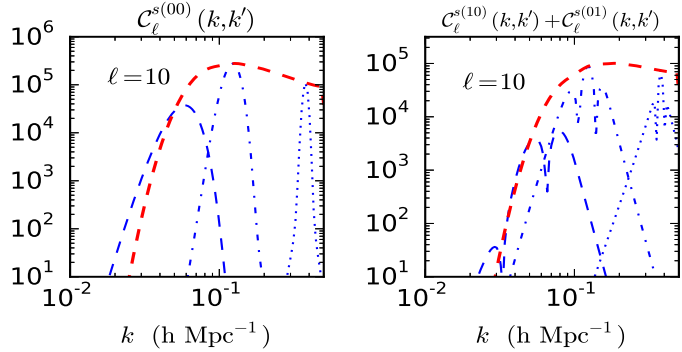}}} 
\end{tabular}
\caption{Same as Figure-\ref{fig:cov2} but for $\ell=10$ and with
$k^{\prime} = 0.05 ({\rm dashed\;lines}), 0.1 ({\rm dotted\; dashed\; lines})$ and $0.35 ({\rm dotted\; lines}) \; h\,\rm{Mpc}^{-1}$ from left to right.}
\label{fig:cov10}
\end{figure} 

The power spectra are diagonally dominated. However, a few off-diagonal terms of the  
matrix $\tilde{\cal C}^s_{\ell}\,(k,k')$ are displayed in Figure-\ref{fig:cov2} (for $\ell=2$) and  
Figure-\ref{fig:cov10} (for $\ell=10$). 
The left and right panels show $\tilde{\cal C}_{\ell}^{s(00)}$ and 
$\tilde{\cal C}_{\ell}^{s(10)}+\tilde{\cal C}_{\ell}^{s(01)}$, respectively. 
For higher $k'$ the covariance matrix is more sharply peaked at $k=k'$. 
The slices through the covariance matrix for $\ell=10$ as depicted in 
Figure-\ref{fig:cov10} are more sharply peaked compared to the $\ell=2$ results for 
the same values of $k'$.
These properties follow from the behaviour of the window functions $W_{\ell}^{s}(k,k')$ displayed
in Figure-\ref{fig:kernel} and -\ref{fig:kernel10}.
To compare the diagonal elements with the off-diagonal terms we have also shown 
$\tilde{\cal C}_\ell^s(k,k)$ in these plots (dashed-lines).

\subsection{Impact of the modified gravity models}
\begin{figure}

\centering
{\epsfxsize=15.5 cm \epsfysize=4.75 cm{\epsfbox[31 425 550 590]{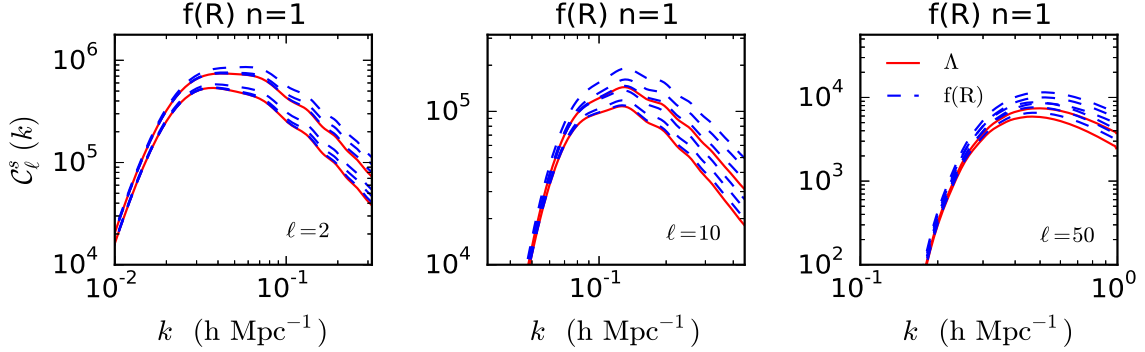}}} 
\caption{The 3D linear power spectra $\tilde{\cal C}^s_{L\ell}(k)$ 
that represent the diagonal elements of the 3D sFB power spectrum $\tilde{\cal C}^s_{L\ell}(k,k)$ as defined in Eq.(\ref{eq:cls_def})
for $f(R)$, $n=1$, theories of gravity are plotted
as a function of the radial wave number $k$ for $\ell=2$ (left-panel), $\ell=10$ (middle-panel) and $\ell=50$ (right-panel), respectively. 
In each panel the dashed curves correspond to the three different parameter values $f_{R_0}=-10^{-4}, -10^{-5}$ 
and $-10^{-6}$ (from top to bottom). The $\Lambda$CDM result is shown with solid curves. 
Each panel displays two different sets of curves. The top and lower sets correspond
to sFB power spectra $\tilde{\cal C}^s_{L\ell}(k)$
with and without the redshift space distortions.}
\label{fig:fofRn1}
\end{figure} 
\begin{figure}
\centering
\vspace{0.5cm}
{\epsfxsize=15.5 cm \epsfysize=4.75 cm{\epsfbox[31 425 550 590]{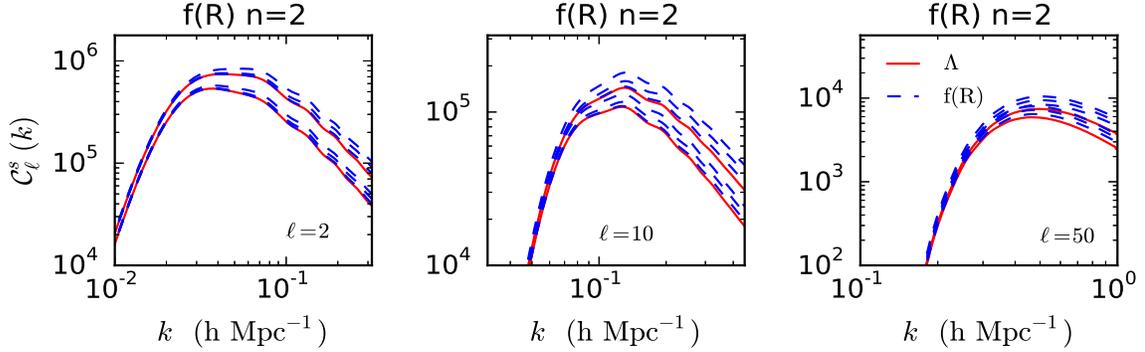}}}
\caption{Same as Figure-\ref{fig:fofRn1} but for $n=2$.}
\label{fig:fofRn2}
\end{figure} 
In Figure-\ref{fig:fofRn1} and Figure-\ref{fig:fofRn2} we plot the diagonal entries of the 
3D sFB power spectrum $\tilde{\cal C}^s_{\ell}(k,k)$, as defined in Eq.(\ref{eq:cls_def}), 
as a function of the wave number $k$ for $f(R)$ gravity theories. 
The results correspond to $\ell=2,10$ and $50$ respectively. 
The range of $k$-values probed is $0.01\,\rm h Mpc^{-1}$-$0.2\,\rm h Mpc^{-1}$.
In Figure-\ref{fig:fofRn1} we show the results for $n=1$ and in Figure-\ref{fig:fofRn2} 
we show the results for $n=2$.
For each value of $n$ we show the three values $|f_{R_0}|=\{10^{-4},10^{-5},10^{-6} \}$.
The base $\Lambda$-CDM model is also plotted (solid lines). 
In comparison to the $\Lambda$-CDM model all $f(R)$ models have additional power 
at all $k$. 
We find that for all values of $k$ and $\ell$ the redshift-space contribution 
${\cal C}^{s(10)}_{\ell}+{\cal C}^{s(01)}_{\ell}$ is positive. 
We will see that the inclusion of redshift information improves our ability to distinguish 
departures of MG theories from GR.
\begin{figure}
\centering
\vspace{0.15cm}
\begin{tabular}{cccc}
{\epsfxsize=15.5 cm \epsfysize=4.75 cm{\epsfbox[30 425 550 590]{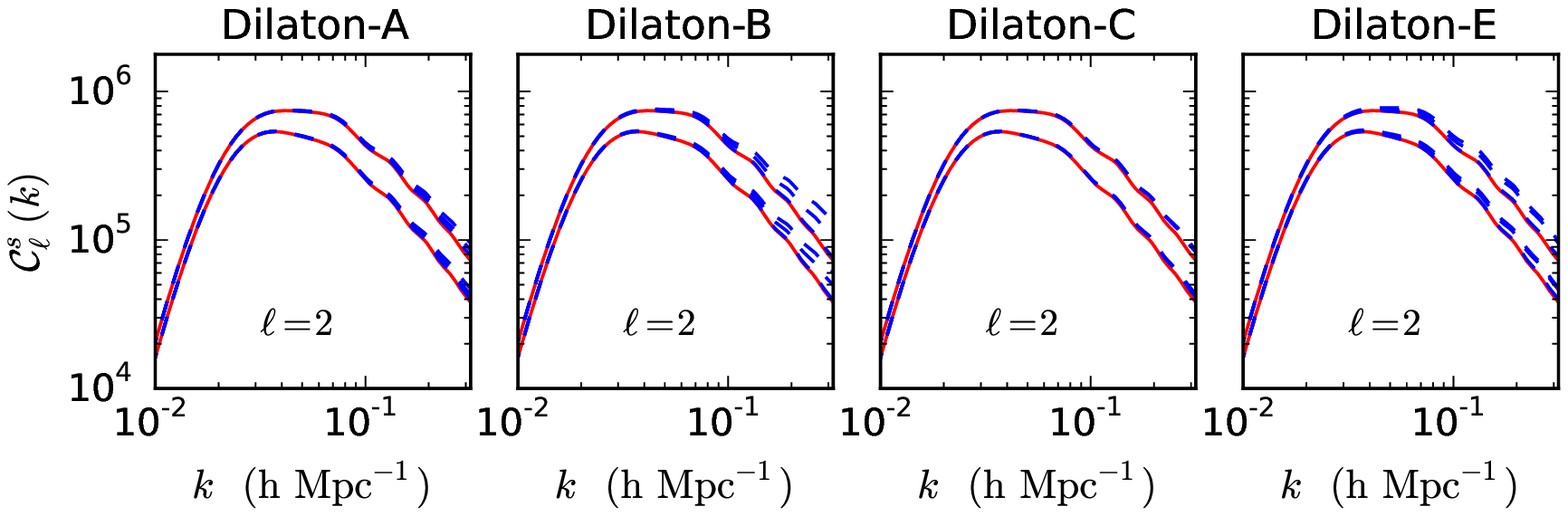}}} 
\end{tabular}
\caption{The 3D linear power spectra $\tilde{\cal C}^s_{L\ell}(k)$, as in Figure~\ref{fig:fofRn1}
but for the Dilaton models, are shown as a function of the radial wave number $k$ for the 
angular harmonic $\ell=2$. In each panel the solid curves represent the $\Lambda$CDM 
result while the dashed curves represent the Dilaton models. 
From left to right the panels correspond
to the Dilaton models A, B, C and E, respectively. The upper (lower) set of curves in each 
panel corresponds to results with (without) redshift space distortions. 
For each chosen Dilaton Gravity model we plot the power spectra for the three choices 
of parameter values given in Table-\ref{tabular:tab2}.
The models A, B, C and E correspond to variation of the parameters 
$s$, $\beta_0$, $r$ and $m_0$ respectively.}
\label{fig:dilaton1}
\end{figure} 
\begin{figure}
\centering
\vspace{0.15cm}
{\epsfxsize=15.5 cm \epsfysize=4.75 cm{\epsfbox[30 425 550 590]{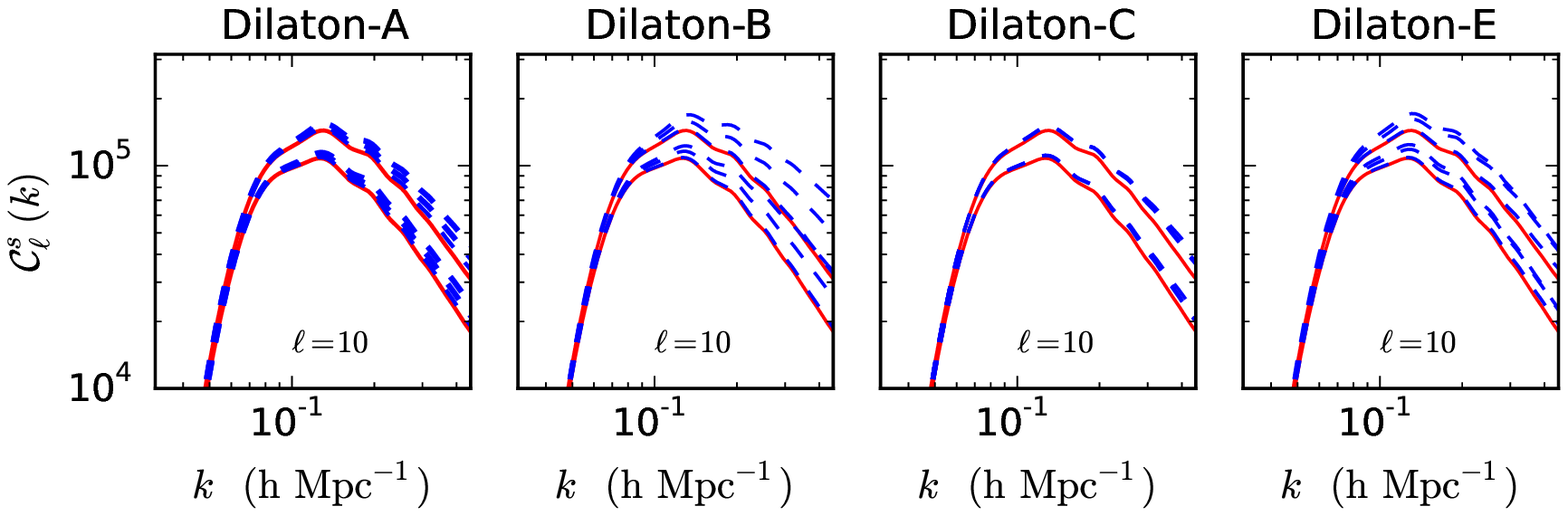}}}
\caption{Same as Figure-\ref{fig:dilaton1} but for $\ell=10$.}
\label{fig:dilaton2}
\end{figure} 
\begin{figure}
\centering
\vspace{0.15cm}
{\epsfxsize=15.5 cm \epsfysize=4.75 cm{\epsfbox[31 425 550 590]{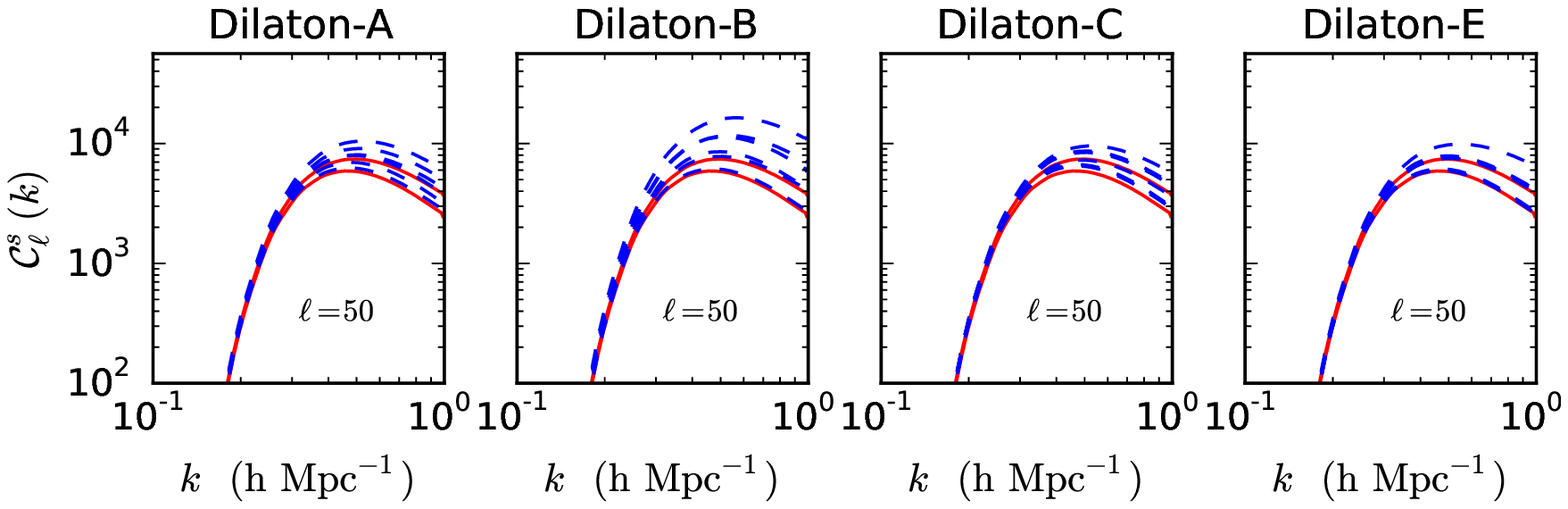}}} 
\caption{Same as Figure-\ref{fig:dilaton1} but for $\ell=50$.}
\label{fig:dilaton3}
\end{figure} 
The Dilaton models we consider are specified by the four parameters  $m_0$, $r$, 
$\beta_0$ and $s$, as given in Table-\ref{tabular:tab2}. 
The results for $\ell=2$, $10$ and $50$ are presented in Figures-\ref{fig:dilaton1},
-\ref{fig:dilaton2} and -\ref{fig:dilaton3}, respectively. 
In agreement with the 3D power spectra shown in Figure-\ref{fig:ps_MG},
over the range $k \leq 1 h \rm Mpc^{-1}$ and $\ell \leq 50$ associated with linear
and weakly non-linear scales, the relative deviations from the $\Lambda$CDM power
grow at higher radial wave number $k$ and angular harmonic $\ell$.
\section{Covariance and $\chi^2$} 
The likelihood function $\cal L$ for arbitrary sets of parameter $\Theta_{\mu}$ 
(that specify a given MG theory; e.g. $\{n,f_{R_0}\}$ in the case of $f(R)$ theory), 
given the data vector $\tilde{\cal C}^s_{\ell}(k)$ (which consists of the {\em noisy} 
3D sFB power-spectra $\tilde{\cal C}^s_{\ell}(k) \equiv \tilde{\cal C}^s_{\ell}(k,k)$), 
is given by:
\ben
&& {\cal L}(\Theta_{\mu} | \tilde{\cal C}^s_{\ell}(k)) = {1 \over (2\pi)^{N_{\rm pix}/2} 
| \det\; {\mathbb C} |^{1/2}} 
\exp \left [-{1 \over 2} \sum_{\ell\ell'}\int dk \int dk' \; \delta{\cal C}^s_{\ell}(k)\; 
{\mathbb C}^{-1}_{\ell\ell'}(k,k')\; \delta{\cal C}^s_{\ell'}(k') \right ] .
\een
Here 
$\delta \tilde{\cal C}^s_{\ell} = \tilde{\cal C}^s_{\ell} - \tilde{\cal C}^{s \rm GR}_{\ell}$ 
(${\cal C}^{s \rm GR}_\ell$ being the power spectrum in GR);  
$N_{\rm pix}$ is the size of the data vector, which depends on the angular resolution 
$\ell_{max}$ and the number of radial bins used in the computation. 
The covariance matrix ${\mathbb C}_{\ell\ell'}(k,k')$ is given by:
\ben
&& {\mathbb C}_{\ell\ell'}(k,k') \equiv \langle {\tilde{\cal C}}^s_{\ell}(k)
{\tilde{\cal C}}^s_{\ell'}(k')\rangle -\langle {\tilde{\cal C}}^s_{\ell}(k)\rangle 
\langle {\tilde{\cal C}}^s_{\ell'}(k')\rangle  = 
{2\over 2\ell +1} \left [\tilde{\cal C}^s_{\ell}(k,k')+ {1\over {\rm \bar N}}  \right ]^2
\delta_{\ell\ell'}\delta_{m\,m'} ,
\label{eq:cov}
\een
where we used a Gaussian approximation. Here $\rm\bar N$ represents the number density of galaxies which are assumed to be 
{\em Poisson} distributed. The block diagonal form is a result of assuming an all-sky coverage. 
However we found the $\mathbb C$ to be diagonally dominant. 
Partial sky coverage will introduce off-diagonal terms between different harmonics 
in the covariance matrix. A rough scaling of signal-to-noise (S/N) with the fraction
of sky-coverage $f_{\rm sky}$ is typically used in the literature: ${\rm S/N}\propto\sqrt{f_{\rm sky}}$. 
We work with the $\chi^2$ statistics defined as:
\ben
&& \chi^2 = \sum_{\ell\ell'}\int dk \int dk' \; \delta{\tilde{\cal C}}^s_{\ell}(k)\; 
{\mathbb C}^{-1}_{\ell\ell'}(k,k')\; \delta{\tilde{\cal C}}^s_{\ell'}(k').
\een
We assume a perfect knowledge of all background cosmological parameters and 
present results with and without RSD.
In practice, we bin the radial wave number $k$ in four logarithmic bins to avoid the 
covariance matrix being singular, thus replacing the integrals over $k$ and $k'$ by 
discrete sums, and we use the multipoles $\ell = 1$ to $25$.
The galaxy number density is fixed at ${\rm \bar N} = 10^{-4} \rm Mpc^{-3}$. 
We again assume all-sky coverage and the survey depth is fixed at 
$r_0=150 h^{-1}\rm Mpc$.
\begin{figure}
\centering
\vspace{0.75cm}
{\epsfxsize=10.15 cm \epsfysize=4.25 cm{\epsfbox[36 408 408 590]{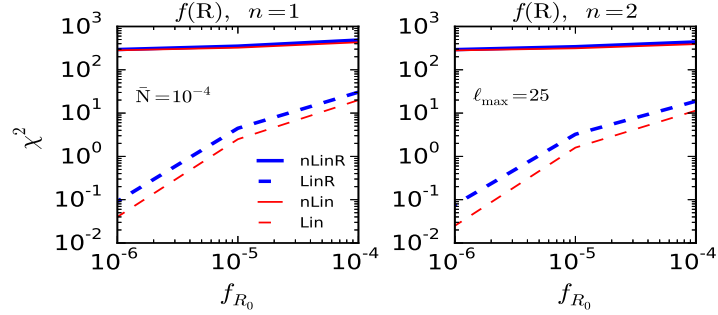}}}
\caption{The $\chi^2$ for the $f(R)$ models are presented as a function of the 
parameter $|f_{R_0}|$. 
The left panel correspond to $n=1$ and the right panel to $n=2$. 
The two solid curves near the top of the panels correspond to 
results using non-linear power spectrum 
(\ref{eq:PowSpec}). The curve with higher $\chi^2$ for a given $f_{R_0}$
includes RSD contribution (nLinR) while the one (nLin) with lower $\chi^2$ doesn't.
The two dashed curves at the bottom of the panel
are the results derived using linear power spectrum (\ref{eq:powUU})
with and without RSD contribution denoted as LinR and Lin respectively. Inclusion of RSD improves
the $\chi^2$ in both linear and non-linear regime.}
\label{fig:chi_fofR}
\end{figure}
We show our results for the $f(R)$ theories in Figure-\ref{fig:chi_fofR}.
For the $n=1$ model we find that for $\bar{\rm N}= 10^{-4}\rm Mpc^{-3}$ the values of 
$|f_{R_0}| > 2 \times 10^{-5}$ can be ruled out with a $3\sigma$ confidence. 
For $n=2$ we find the constraint degrades to $|f_{R_0}| >  3 \times 10^{-5}$.
The $f(R)$ models with higher exponent $n$ are progressively less constrained
as they converge increasingly fast to $\Lambda$CDM with redshift at $z>0$.
\begin{figure}
\centering
\vspace{0.75cm}
{\epsfxsize=18.15 cm \epsfysize=4.75 cm{\epsfbox[30 440 550 600]{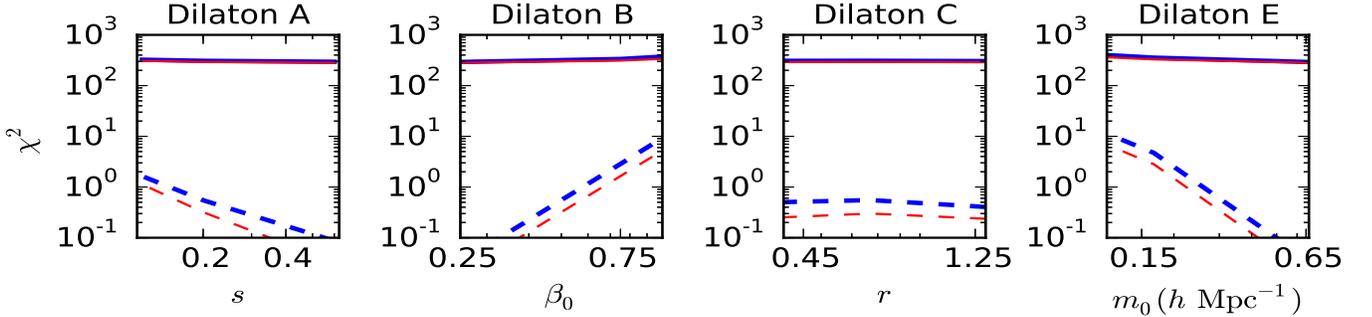}}}
\caption{The $\chi^2$ for various generalised Dilaton models (see Table-\ref{tabular:tab2}) 
are shown. From left to right various panels correspond to models A, B, C and E. 
Each model corresponds to variation of a specific parameter 
among $\{s,\beta_0,r,m_0\}$ while keeping all other parameters fixed. 
The $\chi^2$ in panels from left to right is shown as a function of $s$, $\beta_0$, 
$r$ and $m_0$. The line-style is the same as in Figure-\ref{fig:chi_fofR}.}
\label{fig:Dilaton10chi}
\end{figure}
We show our results for the Dilaton models in Figure-\ref{fig:Dilaton10chi}.
We consider the problem of the estimation of each individual parameter
$\{s,\beta_0,r,m_0\}$ while keeping others fixed, which corresponds to the
model families A, B, C and E.
For $\bar{\rm N} = 10^{-4} \rm Mpc^{-3}$ we find that the parameter values $\beta_0>0.8$ and 
$m_0<0.15 h {\rm Mpc^{-1}}$ can be ruled out with $3\sigma$ confidence. 
No meaningful constraints on $\sigma$  or $r$ can be obtained at $3\sigma$ level.
Clearly, a joint estimation will be more demanding. 
These parameters will also have some degeneracy with the parameters describing 
the background cosmological dynamics. 
A joint Fisher analysis of MG parameters and cosmological parameters will
require a dedicated study and will be presented elsewhere.
As the Planck observations provide an accurate baseline standard cosmological model,
using Planck prior may be a useful practical solution. 

The comparison of the linear and non-linear curves in Figures-\ref{fig:chi_fofR} and
-\ref{fig:Dilaton10chi} suggests that the non-linear contributions can make a significant
effect. However, in the non-linear regime the covariance matrix should include the 
bispectrum and the trispectrum of the density field, which would reduce the $\chi^2$.
Therefore, we can expect that a fully non-linear analysis, with a better modelling of the
covariance matrix (e.g., from numerical simulations) would lower down the non-linear
curves and make them closer to the linear result. However, because the departures
from the $\Lambda$CDM power spectrum increase on weakly non-linear scales, non-linearities
should still improve the discriminatory power of the analysis of 3D clustering as compared
with the linear result.
Here we follow a conservative approach as we estimate the constraints on the modified-gravity
parameters by using the linear-theory $\chi^2$.

We have studied the constraints in both $f(R)$ and Dilaton models with and without 
redshift space distortion. 
We find that redshift space distortions only lead to a small broadening of the
constraints.
Below galaxy number density of ${\rm \bar N}=10^{-4} \rm Mpc^{-3}$ the surveys rapidly lose their ability 
to discriminate. On the other hand, increasing the number density beyond 
${\rm \bar N}= 10^{-3} \rm Mpc^{-3}$ does not lead to drastic improvements of the results.

Our constraints are derived for a spectroscopic survey. 
Inclusion of photometric redshift error will degrade the discriminating power of the 
survey. 

Our constraints are based on lower order tangential modes $\ell \le 25$.
We find that the information contents in different $\ell$ are highly degenerate.
In the range $0.01 \leq k \leq\, 0.2 h\,{\rm Mpc}^{-1}$, 
we find that for more than three to four (logarithmic) bins in $k$ the covariance matrix 
can become singular. 

Finally, our results are based on an uniform distribution of noise (constant ${\rm \bar N}$) 
while real surveys may have more complicated variations in the average galaxy 
number density which may depend on the radial and angular coordinates in the sky. 
\section{Discussion \& Future Prospects}
In this paper we have studied the possibility of constraining modified gravity theories
using galaxy clustering. We have studied two  different modified gravity models:
$f(R)$ theories and {\em Dilaton} theories. In both cases we have used
a specific parametrization and computed the $\chi^2$ as a criterion for 
constraining the departure from the $\Lambda$CDM models. We used the
sFB transforms and the resulting 3D power spectrum ${\cal C}_{\ell}(k)$
for a range of $\ell$ and $k$ values to constrain the model.
We assume an all-sky coverage and a spectroscopic survey with a Gaussian selection function 
$\varphi(r)\propto \exp(-{r^2/r^2_0}), r_0=150h^{-1}${\rm Mpc}. We fix the number density
of galaxies  to be $\bar {\rm N} =10^{-4}\;{\rm Mpc}^{-3}$.
We find that the low $\ell \leq 25$ modes of ${\cal C}^s_\ell(k,k')$ 
(with radial modes restricted to $k<0.2\,h\,{\rm Mpc^{-1}}$) can constraint the 
parameter $f_{R_0}$ at a level of $2\times 10^{-5} (3\times 10^{-5})$ with 
$3 \sigma$ confidence for $n=1(2)$. For the Dilaton models some of the 
parameters ($\beta_0, m_0$) can be well constrained using galaxy clustering though there 
are others ($s,r$) which remain poorly constrained.  The parametrization
used by us depends on a  tomographic approach.
Combining constraints from higher $\ell\ge 25$ modes can further reduce the error-bar 
and thus in principle make cosmological probes of gravity competitive with solar system 
tests. However this will require an accurate modelling of non-linear clustering
in redshift space as well as the covariance matrix of the sFB power spectra.
Our results are based on the linear power spectrum. However, we find that 
inclusion of nonlinear effect can drastically improve the $\chi^2$.
However we would like to emphasise that
the contribution from higher order moments notably the trispectrum
is not included in our covariance matrix. Although, this may not
play an imprortant role in the quasi-linear regime, in the nonlinear regime
such contributions can no longer be ignored. A detailed analysis
will be presented elsewhere. Partial sky coverage, boundary conditions
from specific survey geometry will also mean that in a realistic survey the sFB modes
will have to be discrete and not continuous even in the radial direction.
The average number density of galaxies and hence the noise too will be a function of the angular position
on the surface of the sky. A pseudo-${\cal C}_{\ell}$ (PCL) based approach can be useful
in this direction. The errors due to photometric redshift determination can be readily
included in our calculation. 

We have only considered two sets of MG theories. However, the results derived are quite
general and can readily be extended to other MG scenarios such as the 
K-mouflage models or the Symmetron models. In addition,
the sFB power spectrum can also be useful in constraining massive neutrinos, 
warm dark matter candidates or axionic dark matter through their footprints on the 
matter power spectrum as a function of redshift.

The number of parameters required to specify a MG model can be high.
Non-parametric techniques such as Principal Component Analysis (PCA) 
can be efficient to investigate linear combinations (ordered according to decreasing 
signal-to-noise) of parameters that can be extracted using a specific survey strategy.
Extending the $\chi^2$ based approach presented here, 
a more extensive analysis of survey optimisation
covering a range of survey parameter and associated Fisher matrix analysis detailing
parameter degeneracies will be presented elsewhere. In future we also plan to use
a Bayesian model selection approach based on the evidences for acceptance or rejection of 
specific models.  

Physics of galaxy formation may be different in modified gravity models.
Even in the $\Lambda$CDM cosmology, the bias associated with the galaxies may 
depend on the galaxy type and their environment, and may be stochastic. 
Such complications can only be dealt with numerical simulations. 
However, one may expect that at very large scale such effects may be less important.

CMB mostly probes high redshifts where any modification from GR is expected to be 
sub-dominant. However, CMB is also sensitive to modifications of gravity through secondary effects
such as the Integrated Sachs-Wolfe (ISW) effect and the weak lensing distortions 
\citep{PC15}. However the constraints are rather weak compared to what can be achieved by 
galaxy surveys \citep{MuHu14}. 

Weak lensing is a very exciting possibility for constraining any departure 
from GR especially because it is free from the problem of galaxy bias \citep{MVWH08}.
However, in recent years many possible systematics have been
discussed, e.g. from intrinsic ellipticity correlations or 
possible modifications of the power spectrum due to AGN feedback, gas cooling and 
stellar feedback or neutrino physics (Munshi et al. 2015; in preparation).
Such contaminations may induce appreciable bias in parameter estimation 
(e.g. \citep{OSY15}).
Despite all these systematics, weak lensing and galaxy clustering remain the most 
powerful probe of possible departures from GR on cosmological scales.

Statistics of Lyman-$\alpha$ absorbtion have also been investigated in the context of 
$f(R)$ gravity \citep{APV15}.
Using cosmological hydrodynamical simulations of $f(R)$ gravity which include the flux 
probability distribution functions and the flux power-spectra, and an analysis of the column 
density and line-width distributions, as well as the matter power spectrum,
It was found that Lyman-$\alpha$ statistics is rather insensitive to modification to gravity in 
$f(R)$ models.
Hence no competitive constraints are achievable using current data. 
Moreover, contamination by baryonic physics, associated with star formation and 
cooling processes, imply that a very accurate modelling of such ingredients is required. Constraints from $21$cm intensity mapping are more encouraging \citep{Hall13}.

Observations by SDSS galaxy surveys in the low to moderate redshift range 
$0.15 < z<0.67$  is already being used to set interesting constraints on any departure 
from GR, e.g. \cite{Julien14} obtained $f_{R_0}<4.6\times 10^{-5}$ at the 
$95\%$ confidence level. 
Euclid with survey characteristics similar to SDSS will probe
deeper and wider parts of the sky thus improving the constraints by orders of magnitude.

\section*{Acknowledgements}
D.M. and P.C. acknowledge support from the Science and Technology Facilities Council 
(grant number ST/L000652/1). The research leading to these results has received funding from the European Research Council under 
the European Union’s Seventh Framework Programme (FP/2007-2013) / ERC Grant Agreement No. [616170].
P.V. acknowledges support from the French Agence Nationale de la Recherche under Grant ANR-12-BS05-0002.
D.M. would like to thank A. Starobinsky and A. Heavens for useful discussions. 
It is a pleasure for D.M. to acknowledge related collaborations with B. Hu, L. van Waerbeke and J. Harnois-Deraps.
\bibliography{red.bbl}
\end{document}